\documentclass{sig-alternate}

\usepackage{times}
\usepackage{subfigure}
\usepackage{url}
\usepackage{multirow}
\usepackage{amssymb,amsmath}
\usepackage{graphicx,afterpage}
\usepackage{algpseudocode}
\usepackage{comment}
\usepackage{mdwlist}
\usepackage[dvips,colorlinks=true,citecolor=black,linkcolor=black,urlcolor=black]{hyperref}
\usepackage[sort]{cite}
\usepackage[hyphenbreaks]{breakurl}
\usepackage{xspace}

\newcommand{\sys}{NetFence\xspace}
\newcommand{\eg}{\emph{e.g.}}
\newcommand{\ie}{\emph{i.e.}}
\newcommand{\bad}{$mon$\xspace}
\newcommand{\ok}{$nop$\xspace}
\newcommand{\hi}{$decr$\xspace}
\newcommand{\lo}{$incr$\xspace}
\newcommand{\ldown}{$L^\downarrow$\xspace}
\newcommand{\lup}{$L^\uparrow$\xspace}
\newcommand{\fp}{\vspace*{0.05in}\noindent}
\renewcommand{\paragraph}[1]{\fp {\bf #1}~}

\title{\sys: Preventing Internet Denial of Service from Inside Out}
\subtitle{Technical Report 2010-01}

\numberofauthors{3}
\author{
\alignauthor
Xin Liu \\
\affaddr{Dept. of Computer Science} \\
\affaddr{Duke University} \\
\email{xinl@cs.duke.edu}
\alignauthor
Xiaowei Yang \\
\affaddr{Dept. of Computer Science} \\
\affaddr{Duke University} \\
\email{xwy@cs.duke.edu}
\alignauthor
Yong Xia \\
\affaddr{Networking Systems Group} \\
\affaddr{NEC Labs China} \\
\email{xia\_yong@nec.cn}
}

\begin{document}

\maketitle
\begin{abstract}

Denial of Service (DoS) attacks frequently happen on the Internet,
paralyzing Internet services and causing millions of dollars of
financial loss. This work presents NetFence, a scalable DoS-resistant
network architecture. NetFence uses a novel mechanism, secure congestion
policing feedback, to enable robust congestion policing inside the
network. Bottleneck routers update the feedback in packet headers to
signal congestion, and access routers use it to police senders'
traffic. Targeted DoS victims can use the secure congestion policing
feedback as capability tokens to suppress unwanted traffic. When
compromised senders and receivers organize into pairs to congest a
network link, NetFence provably guarantees a legitimate sender its fair
share of network resources without keeping per-host state at the
congested link. We use a Linux implementation, ns-2 simulations, and
theoretical analysis to show that NetFence is an effective and
scalable DoS solution: it reduces the amount of state maintained by a
congested router from per-host to at most per-(Autonomous System).

\end{abstract}

\section{Introduction}
\label{sec:intro}

Large-scale Denial of Service (DoS) attacks remain as a potent threat
to the Internet. A survey from Arbor Networks shows that DoS attacks
continue to grow in both scale and
sophistication~\cite{arbor-report-5}. The largest observed attack
reached 49Gbps in 2009, a 104\% growth over the past two years.  The
survey also ranks DoS attacks as the largest anticipated threat in the
next 12 months. This result is not surprising, as tens of gigabits
flooding traffic could easily overwhelm most links, routers, or sites
on the Internet.

The destructive nature of DoS attacks has brought forth a fundamental
research challenge: how can we design an open network architecture
that is resistant to large-scale DoS attacks?  There have been several
proposals addressing this
challenge~\cite{aitf-ton,stopit,portcullis,tva-ton,siff,anderson-hotnets-2003}.
These proposals enable DoS victims to suppress attack traffic using
network capabilities or filters, but when malicious sender-receiver
pairs collude to flood a link, the best defense mechanism these
systems can offer is per-host queuing at the flooded link to separate
legitimate traffic from attack traffic. This solution faces a
scalability challenge, as a flooded router may forward packets for
millions of (malicious and legitimate) end systems.

This paper presents the design and evaluation of \sys, a scalable
DoS-resistant network architecture. \sys provably guarantees each
sender its fair share of bandwidth without keeping per-host state at
bottleneck routers even when malicious senders and receivers collude
into pairs to flood the network.  It also enables DoS victims to
suppress unwanted traffic as in a capability-based
system~\cite{tva-ton,portcullis}. A key departure of \sys from
previous work is that it places the network at the first line of DoS
defense rather than relies on end systems (be it senders or receivers)
to suppress attack traffic.

The \sys design places a robust traffic policing control loop inside
the network (\S~\ref{sec:arch} and \S~\ref{sec:design}). Packets carry
unforgeable congestion policing feedback stamped by routers that
suffer excessive congestion (caused either by DoS attacks or other
reasons, which \sys does not distinguish). Access routers at the trust
boundaries between the network and end systems examine the feedback
and police the senders' traffic. A malicious sender cannot gain more
than its fair share of bandwidth even if it colludes with a
compromised receiver, because it cannot spoof valid congestion
policing feedback. Innocent DoS victims can use the unforgeable
congestion policing feedback as capability tokens to suppress the bulk
of unwanted traffic, by not returning the feedback to malicious
senders. To be fail-safe in case access routers are compromised, \sys
uses Autonomous System (AS)-level queues (or rate-limiters) to
separate traffic from different source ASes, limiting DoS damage to
the ASes that harbor the compromised routers.

We have implemented \sys in Linux and evaluated its overhead and
performance using theoretical analysis (\S~\ref{sec:arch_theorem}),
testbed experiments, and large-scale simulations
(\S~\ref{sec:eval}). Our analysis shows that regardless of attackers'
strategies, \sys provides a legitimate sender its fair share of
bottleneck bandwidth. The simulation results correlate well with this
analysis, and also show that \sys performs similarly to
state-of-the-art capability- or filter-plus-fair-queuing DoS defense
systems~\cite{stopit,tva-ton}. Our Linux prototype benchmarking
results show that \sys's per-packet processing overhead is low.

These results suggest that \sys is an effective and scalable DoS
solution. \sys's bottleneck routers have $O(1)$ per-packet
computational overhead, and maintain at most per-AS state (more
scalable design alternatives exist as discussed in
\S~\ref{sec:localize_damage}), while previous work requires these
bottleneck routers to keep per-host state to protect legitimate
traffic. One concern for the \sys design is that access routers need
to keep per-(sender, bottleneck link) state (\S~\ref{sec:arch}), but we
show in \S~\ref{sec:scalability}
today's access routers can meet such scalability requirements.

The key contributions of this paper include a new DoS defense
primitive: secure congestion policing feedback, and based on it, the
construction of a robust, network-based, closed-loop congestion
policing architecture that scalably and effectively limits the damage
of DoS flooding attacks. With a closed-loop design, \sys can flexibly
place different functionalities at different locations: lightweight
attack detection and congestion signaling at bottleneck links, and
congestion policing that requires per-(sender, bottleneck link) state
at access routers. This design makes it scale much better than
previous open-loop approaches that employ per-host queuing at
bottleneck routers~\cite{stopit,tva-ton}.

\section{Assumptions and Goals}
\label{sec:goals}

 Before we present the design of \sys, we first describe its threat
 model, assumptions, and design goals.

\subsection{Threat Model and Assumptions}

\paragraph{Flood-based network attacks:} \sys focuses on mitigating
network-layer flooding attacks where attackers send excessive traffic
to exhaust network resources such as link capacity or router
processing power. It does not aim to mitigate DoS attacks that exploit
application vulnerabilities to exhaust end system resources.

\paragraph{Strong adversary:} We assume that attackers can compromise
both end systems and routers. Compromised end systems involved in an
attack can grow into millions; they may launch brute-force or
strategic flooding attacks. For instance, they may disguise attack
traffic as legitimate traffic, launch on-off attacks, or collude into
sender-receiver pairs to send flooding traffic. Attack traffic may or
may not be distinguishable from legitimate traffic.

We make two assumptions to assist \sys's design.

\paragraph{Trust:} We assume that routers managed by the network are
much less likely to be compromised than end systems. We thus place
policing functions on routers rather than end systems. As a tradeoff
for scalability, we treat each AS as a trust and fate sharing unit.
When compromised routers exist, we aim to localize the damage to the
ASes that harbor compromised routers rather than protect all the
legitimate hosts within such ASes.

\paragraph{Line-speed lightweight cryptography:} We assume that
symmetric key cryptography can be supported at line-speed. Some
current hardware can support AES operations at
40Gbps~\cite{aes-hardware}, and the latest Intel Westmere processors
have native support for AES~\cite{intel-aes}.

\subsection{Goals}

\sys aims to meet several design goals. It is these goals that
distinguish \sys from previous work.

\paragraph{i) Guaranteed network resource fair share:} When DoS
victims can identify attack traffic, we aim to enable them to suppress
the attack traffic near the origins. This prevents attack traffic from
wasting network resources. When DoS victims fail to identify attack
traffic, or when attackers collude into sender-receiver pairs to flood the
network, we resort to a weaker goal to guarantee a legitimate sender
its fair share of network resources. That is, for any link of capacity
$C$ shared by $N$ (legitimate and malicious) senders, each sender with
sufficient demand should be guaranteed at least $O(\frac{C}{N})$
bandwidth share from that link.  This mitigates the effect of
large-scale DoS attacks from denial of service to predictable delay of
service.

\paragraph{ii) Open network:} \sys aims to keep the network open to
new applications, and thus places the attack traffic identification
function at the receivers to avoid false positives introduced by
in-network traffic classification. This goal is also shared by
previous work~\cite{anderson-hotnets-2003,tva-ton,aitf-ton}.

\paragraph{iii) Scalable and lightweight:} \sys may face millions of
attackers that attempt to congest a single link in the network. To be effective at
such a scale, it does not assume that a router always has sufficient
resources to warrant per-flow or per-host state management. It aims to
keep little or no state in the core network and avoid heavyweight
operations such as per-flow/host fair queuing in the core network. To
facilitate high-speed router implementation, \sys aims to incur low
communication, computation, and memory overhead.

\paragraph{iv) Robust:} \sys\ should be robust against both simple,
brute-force flooding attacks and sophisticated ones that attempt to
bypass or abuse \sys\ itself.

\paragraph{v) Incrementally adoptable:} We aim to make \sys
incrementally deployable on today's Internet. Specifically, we aim to
provide early adopters immediate deployment benefits: they can form an
``overlay'' network of deployed regions and benefit collectively from
the deployment. We aim not to require hop-by-hop deployment from a
congested link to compromised end systems to be effective,
unlike~\cite{pushback}.

\paragraph{vi) Network self-reliant defense:} We aim for a
self-reliant solution that depends on only routers in the network, not
other infrastructures such as trusted host hardware~\cite{aip} or DNS
extensions~\cite{portcullis}. Our hypothesis is that extra
dependencies increase security risk and may create deployment
deadlocks. That is, without the deployment or upgrade of other
infrastructures, the design is not effective. Hence, there is little
incentive to deploy it, and vice versa.

\section{Architecture}
\label{sec:arch}

\begin{figure}[t]
\centering
\includegraphics[width=\columnwidth]{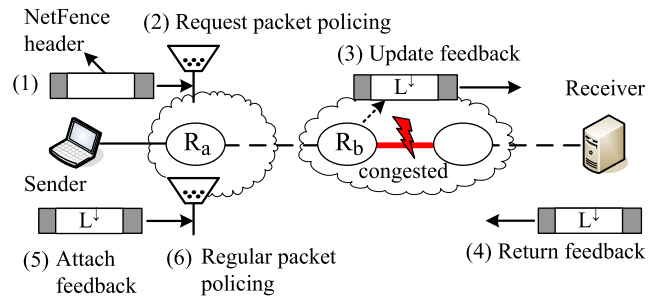}
\caption{\label{fig:overview}{\bf \small The \sys\ architecture.
    Packets carry unspoofable congestion policing feedback stamped by
    bottleneck routers ($R_b$ in this figure). Access routers ($R_a$)
    use the feedback to police senders' traffic, preventing
    malicious senders from gaining unfair shares of bottleneck
    capacity. DoS victims can use the congestion policing feedback as
    capability tokens to suppress unwanted traffic.}}
\end{figure}

In this section, we present an overview of the NetFence architecture,
and defer design details to \S~\ref{sec:design}.

\subsection{System Components}
\label{sec:arch_components}

NetFence has three types of packets: {\em request} packets, {\em
  regular} packets, and {\em legacy} packets. The first two,
identified by a special protocol number in the IP header, have a shim
NetFence header between their IP and upper-layer protocol headers. The
NetFence header carries unforgeable congestion policing feedback
generated by the network (\S~\ref{sec:arch_unforgeable} and
\S~\ref{sec:secure_feedback}). A NetFence-ready sender sends request
and regular packets, while a non-NetFence sender sends only legacy
packets.

\begin{figure}[t]
\centering
\includegraphics[width=\columnwidth]{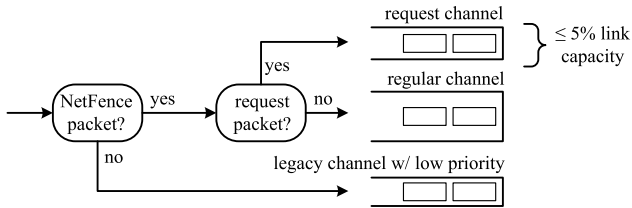}
\caption{\label{fig:queueing}{\bf \small Each \sys\ router keeps three channels.}}
\end{figure}

Each NetFence router, depicted in Figure~\ref{fig:queueing}, keeps
three channels, one for each of the three packet types discussed
above. To motivate end systems to upgrade, the \sys design gives
legacy channel lower forwarding priority than the other two. To
prevent request flooding attacks from denying legitimate requests,
\sys has a priority-based backoff mechanism for the request channel
(\S~\ref{sec:init-pkts}). The request channel is also limited to
consume no more than a small fraction (5\%) of the output link
capacity, as in~\cite{tva-ton,portcullis}.

NetFence places its feedback and policing functions at bottleneck and
access routers that are either inside the network or at the trust
boundaries between the network and end systems. It does not place any
trusted function at end systems. As shown in
Figure~\ref{fig:overview}, a NetFence sender starts an end-to-end
communication by sending request packets to its NetFence-ready
receiver (Step 1). The access router inserts the \ok
feedback in the NetFence header of the packet (Step 2,
\S~\ref{sec:feedback}). Along the path, a bottleneck router might
modify the feedback, in a way similar to TCP ECN~\cite{rfc3168} (Step
3). After the receiver returns the feedback to the sender (Step 4),
the sender can send valid regular packets that contain the feedback (Step
5). In Step 4, two-way protocols like TCP can piggyback the returned
feedback in their data packets, while one-way transport protocols such
as UDP must send extra, low-rate feedback packets from a receiver to a
sender.

A NetFence router periodically examines each output link to decide if
an attack is happening at the link. It uses a combination of link load
and packet loss rate as an attack indicator (\S~\ref{sec:monitoring}).
If an attack is detected, NetFence starts a monitoring cycle, which
lasts until i) no more attack is detected during the cycle, and ii)
the cycle has existed for an extended period (typically a few hours)
after the most recent attack is detected. During a monitoring cycle,
the \bad congestion policing feedback (containing the link ID $l$, an $action$
field, etc.) is stamped into the NetFence header of all the passing
request/regular packets (\S~\ref{sec:update-feedback}). The sender's
regular packets must include this \bad feedback to be considered valid,
and they will be policed by the access router (Step 6,
\S~\ref{sec:police-regular}).

An access router maintains one rate limiter for every
sender-bottleneck pair to limit a sender's regular traffic traversing
a bottleneck link. The router uses an Additive Increase and
Multiplicative Decrease (AIMD) algorithm to control the rate limit: it
keeps the rate limit constant within one pre-defined control interval
(a few seconds); across control intervals, it either increases the
rate limit additively or decreases it multiplicatively, depending on
the particular \bad feedback it receives
(\S~\ref{sec:robust-rate-probing}). We use AIMD to control the rate
limit because it has long been shown to converge onto efficiency and
fairness~\cite{aimd}. Other design choices exist; they have different
cost-performance tradeoffs, and are discussed in
\S~\ref{sec:discussion}.

When no attack is detected, a downstream router will not modify the
\ok feedback stamped by an access router. When the sender obtains the
\ok feedback and presents it back to its access router in a packet,
the packet will not be rate-limited. That is, when no attack happens,
NetFence stays in idle state. The overhead during such idle periods is
low, because 1) the NetFence header is short (20 bytes)
(\S~\ref{sec:pkthdr}); 2) the bottleneck attack detection mechanism
only involves a packet counter and a queue sampler; and 3) an access
router only needs to stamp and validate (not rate limit) the NetFence
header for each packet. Only when an attack is detected at a
bottleneck link, does NetFence activate its policing functions, which
add additional processing overhead at bottleneck and access routers.
We show the overhead benchmarking results in
\S~\ref{sec:benchmarking}.

\subsection{Unforgeable Congestion Policing Feedback}
\label{sec:arch_unforgeable}

Congestion policing feedback must be made unforgeable so that
malicious nodes cannot evade \sys's traffic policing functions. \sys
achieves this goal using efficient symmetric key cryptography.  An
access router inserts a periodically changing secret in a packet's
NetFence header.  A bottleneck router uses this secret to protect its
congestion policing feedback, and then erases the secret. The access router,
knowing the secret, can validate the returned feedback. We describe
the details of this design in \S~\ref{sec:secure_feedback}, and
discuss how to limit the effect of compromised access routers in
\S~\ref{sec:localize_damage}.

\subsection{Congestion Feedback as Capability}
\label{sec:arch_capability}

If a DoS victim can identify and desires to bar attack traffic, \sys's
unspoofable congestion policing feedback also serves as a capability
token: a receiver can return no feedback to a malicious
sender. Because the malicious sender cannot forge valid feedback,
it cannot send valid regular packets. It can at most
flood request packets to a destination, but an access router will use
a priority-based policing scheme to strictly limit a sender's request
traffic rate (\S~\ref{sec:init-pkts}). Alternatively, it can simply
flood to congest its local area network, but this attack is easy to
detect and the damage is confined to the local area network.

\subsection{Fair Share Guarantee}
\label{sec:arch_theorem}

With the above-described closed-loop network architecture, we are able to
prove that NetFence achieves per-sender fairness for single bottleneck
scenarios:

\vspace{0.05in}

{\em \textbf{Theorem}: Given $G$ legitimate and $B$ malicious senders
  sharing a bottleneck link of capacity $C$, regardless of the attack
  strategies, any legitimate sender $g$ with sufficient demand
  eventually obtains a capacity fair share $\frac{\nu_g \rho
    \ C}{G+B}$, where $0 < \nu_g \leq 1$ is a parameter determined by
  how efficient the sender $g$'s transport protocol (\eg, TCP)
  utilizes the rate limit allocated to it, and $\rho$ is a parameter
  close to 1, determined by \sys's implementation-dependent AIMD and
  attack detection parameters.}

\vspace{0.05in} We briefly describe why this theorem holds, but leave
a detailed proof in Appendix~\ref{sec:appendix_proof}.

\textbf{Proof sketch:} In \sys, an access router keeps one rate
limiter for each sender-bottleneck pair when a monitoring cycle is
triggered during attack times.  Based on the unspoofable congestion
feedback from the bottleneck, the access router dynamically adjusts
the rate limits using a robust AIMD algorithm
(\S~\ref{sec:robust-rate-probing}). Since AIMD has been shown to
converge onto efficiency and fairness~\cite{aimd}, all the rate limits
will eventually converge to the fair share of the bottleneck capacity.
Thus, any sender, whether legitimate or malicious, can send at most as
fast as its fair share rate.

In the next section, we will show the design details that make
NetFence achieve this lower bound on fairness despite various
brute-force and strategic flooding attacks.

\section{Design Details}
\label{sec:design}

In this section, we show the design details of \sys. For clarity, we
first present the design assuming unforgeable congestion policing
feedback and non-compromised routers.  We then describe how to make
congestion policing feedback unforgeable and how to handle compromised
routers. Key notations used to describe the design are summarized in
Figure~\ref{fig:parameter}.

\subsection{Congestion Policing Feedback}
\label{sec:feedback}
\sys uses three types of congestion policing feedback:
\begin{itemize*}
\vspace{-0.02in}
\item \ok, indicating no policing action is needed;
\item \ldown, indicating the link $L$ is overloaded, and an access
  router should reduce the traffic traversing $L$;
\item \lup, indicating the link $L$ is underloaded, and an access router can allow more traffic traversing $L$.
\vspace{-0.02in}
\end{itemize*}

We refer to \lup and \ldown as the \bad feedback. Each congestion
policing feedback includes a timestamp to indicate its freshness.

\subsection{Protecting the Request Channel}
\label{sec:init-pkts}

Attackers may simply flood request packets to congest downstream
links. \sys mitigates this attack with two mechanisms. First, it
limits the request channel on any link to a small fraction (5\%) of
the link's capacity, as in~\cite{tva-ton,portcullis}. This prevents
request packets from starving regular packets. Second, it combines
packet prioritization and priority-based rate limiting to ensure that
a legitimate sender can always successfully transmit a request packet
if it waits long enough to send the packet with high priority. This
mechanism ensures that a legitimate user can obtain the valid congestion
policing feedback needed for sending regular packets.

In \sys, a sender can assign different priority levels to its request
packets. Routers forward a level-$k$ packet with higher priority than
lower-level packets, but the sender is limited to send level-$k$
packets at half of the rate of level-($k$-1) packets.  An access
router installs per-sender token-based rate limiters to impose this
rate limit. It removes $2^{k-1}$ tokens from a request packet rate
limiter when admitting a level-$k$ packet. Level-0 packets are not
rate-limited, but they have the lowest priority.

This request channel policing algorithm guarantees that a legitimate
sender can eventually send a request packet to a receiver regardless
of the number of attackers~\cite{portcullis}. It holds because the
arrival rate of request packets decreases exponentially as their
priority level increases. Thus, the arrival rate of high priority
request packets will eventually be smaller than the request channel
capacity.

\sys does not use computational puzzles as in~\cite{portcullis}. This
is because computational resources may be scarce~\cite{lazy-susan}, especially in busy servers and handheld
devices. In addition, \sys's design has the flexibility that an access
router can configure different token refill rates for different hosts
on its subnet. Legitimate servers could be given a higher rate to send
more high priority request packets without purchasing additional CPU
power.

When an access router forwards a request packet to the next hop, it
stamps the \ok feedback into the packet, ensuring that a sender can
obtain valid feedback if the receiver desires to receive from it.

\subsection{Protecting the Regular Channel}
\label{sec:regular-pkts}

Malicious senders may flood regular packets when they can obtain valid
congestion policing feedback from their colluding receivers. We
describe how to mitigate this attack.

\begin{figure}[t]
\centering \small
\begin{tabular}{r|l|l}
Name & Value & Meaning \\ \hline
$l_1$ & 1 ms &  level-1 request packet rate limit \\ \hline
$I_{lim}$ & 2 s & Rate limiter ctrl interval length \\ \hline
$w$ & 4 s & Feedback expiration time \\ \hline
$\Delta$ & 12 kbps & Rate limiter additive incr \\ \hline
$\delta$ & 0.1 & Rate limiter multiplicative decr \\ \hline
$p_{th}$ & 2\% &  Packet loss rate threshold\\ \hline
$Q_{lim}$ & 0.2s $\times$ link bw & Max queue length \\ \hline
$minthresh$ & $0.5 \, Q_{lim}$ & RED algorithm parameter\\ \hline
$maxthresh$ & $0.75 \, Q_{lim}$ & RED algorithm parameter \\ \hline
$w_q$ & 0.1 & EWMA weight for avg queue length
\end{tabular}
\caption{\label{fig:parameter} {\bf \small Key parameters and their
    values in our implementation.}}
\end{figure}

\subsubsection{A Monitoring Cycle}
\label{sec:monitoring}

When a router suspects that its outgoing link $L$ is under attack, it
starts a monitoring cycle for $L$. That is, it marks $L$ as in the
\bad state and starts updating the congestion policing feedback
in packets that traverse $L$ (\S~\ref{sec:update-feedback}). Once a
sender's access router receives such feedback, it will start rate
limiting the sender's regular packets that will traverse the link $L$
(\S~\ref{sec:police-regular}).

It is difficult to detect if $L$ is under an attack because the attack
traffic may be indistinguishable from legitimate traffic. In \sys,
$L$'s router infers an attack based on $L$'s utilization and the loss
rate of regular packets. If $L$ is well-provisioned and its normal
utilization is low (a common case in practice), it can be considered
as under an attack when its average utilization becomes high (\eg,
95\%); if $L$ always operates at or near full capacity, its router can
infer an attack when the regular packets' average loss rate $p$
exceeds a threshold $p_{th}$.  A link's average utilization and $p$
can be calculated using the standard Exponentially Weighted Moving
Average (EWMA) algorithm~\cite{red}. The threshold $p_{th}$ is a local
policy decision of $L$'s router, but it should be sufficiently small
so that loss-sensitive protocols such as TCP can function well when no
attack is detected. Attackers may launch a mild attack and evade the
detection by keeping $p$ below $p_{th}$, but the damage is also
limited.

When the attack detection is based on the packet loss rate $p$, a
flash crowd may also be considered as an attack. We do not distinguish
these two because it is too difficult to do so. As shown by our
simulation results (\S~\ref{sec:eval}), starting a monitoring cycle
for a link does not have much negative impact on a legitimate sender.

It is undesirable to infinitely keep a monitoring cycle due to the
added overhead. Thus, a \sys router terminates a link $L$'s monitoring
cycle when $L$ is no longer under attack (\eg, $p<p_{th}$) for a
sufficiently long period of time $T_b$. The router will mark $L$ as in
the \ok state and stop updating the congestion policing feedback
in packets traversing $L$.  Similarly, an access router will
terminate a rate limiter $(src,L)$ if it has not received any packet
with the \ldown feedback and the rate limiter has not discarded any
packet for $T_a$ seconds.

Routers should set $T_a$ and $T_b$ to be significantly longer (\eg, a
few hours) than the time it takes to detect an attack ($T_d$). This is
because attackers may flood the network again after $T_a$ (or $T_b$)
seconds. By increasing the ratio of the monitored period $min(T_a,
T_b)$ to the unprotected period $T_d$, we reduce the network
disruption time. Network disruption during an attack detection period
cannot be eliminated unless compromised senders are patched up, but we
do not assume routers have this ability.

\subsubsection{Updating Congestion Policing Feedback}
\label{sec:update-feedback}

When a link $L$ is in the \bad state, its router $R_b$ uses the
following ordered rules to update the congestion policing feedback in
any request/regular packet traversing $L$:
\begin{enumerate*}
\vspace{-0.03in}
\item If the packet carries \ok, stamp \ldown;
\item Otherwise, if the packet carries $L'^\downarrow$ stamped by an upstream link $L'$, do nothing;
\item Otherwise, if $L$ is overloaded, stamp \ldown.
\vspace{-0.03in}
\end{enumerate*}

The router $R_b$ never stamps the \lup\ feedback. As we will see in
\S~\ref{sec:police-regular}, only an access router stamps \lup\ when
forwarding a packet. If the \lup\ feedback reaches the receiver of the
packet, it indicates that the link $L$ is not overloaded, because
otherwise the router $R_b$ would replace the \lup\ feedback with the
\ldown\ feedback.

A packet may cross multiple links in the \bad\ state. The access
router must ensure that the sender's rate does not exceed its
legitimate share at any of these links. The second rule above allows
\sys to achieve this goal, gradually. This is because the first link
$L_1$ on a packet's forwarding path that is both overloaded and in the
\bad\ state can always stamp the $L_1^\downarrow$ feedback, and
downstream links will not overwrite it. When the $L_1^\downarrow$
feedback is presented to an access router, the router will reduce the
sender's rate limit for the link $L_1$ until $L_1$ is not overloaded
and does not stamp $L_1^\downarrow$. This would enable the next link
($L_2$) on the path that is both in the \bad state and overloaded to
stamp $L_2^\downarrow$ into the packets. Gradually, a sender's rate
will be limited such that it does not exceed its fair share on any of
the on-path links in the \bad state.

\subsubsection{Regular Packet Policing at Access Routers}
\label{sec:police-regular}

A sender $src$'s access router polices the sender's regular packets
based on the congestion policing feedback in its packets. If a packet
carries the \ok feedback, indicating no downstream links require
congestion policing, the packet will not be rate-limited. Otherwise,
if it carries \lup\ or \ldown, it must pass the rate limiter
$(src,L)$.

We implement a rate limiter as a queue whose de-queuing rate is the
rate limit, similar to a leaky
bucket~\cite{turner86:_new_direc_commun_or_which}. We use the queue to
absorb traffic bursts so that bursty protocols such as TCP can
function well with the rate limiter. We do not use a token bucket
because it allows a sender to send short bursts at a speed exceeding
its rate limit. Strategic attackers may synchronize their bursts to
temporarily congest a link, leading to successful on-off attacks.

When an access router forwards a regular packet to the next hop, it
resets the congestion policing feedback. If the old feedback is
\ok, the access router refreshes the timestamp of the feedback.  If
the old feedback is \ldown or \lup, the access router resets it to
\lup. This design reduces the computational overhead at the link $L$'s
router, as it does not update a packet's feedback if $L$ is not
overloaded.

For simplicity, \sys uses at most one rate limiter to police a regular
packet. One potential problem is that a flow may switch between
multiple rate limiters when its bottleneck link changes. We discuss
this issue in \S~\ref{sec:multi-ratelimiters}.

\subsubsection{Robust Rate Limit Adjustment}
\label{sec:robust-rate-probing}

\begin{figure}[t]
\centering
\includegraphics[width=\columnwidth]{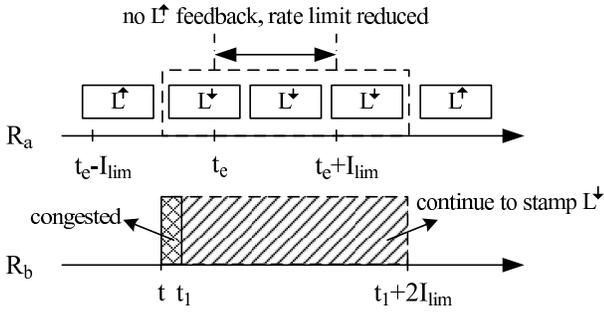}
\caption{\label{fig:robust-signal}{\bf \small Once a router $R_b$
    encounters congestion between time [$t, t_1$], it will
    continuously stamp the \ldown feedback until $t_1+2I_{lim}$.}}
\end{figure}

The \lup and \ldown feedback enables an access router to adjust a rate
limiter $(src, L)$'s rate limit $r_{lim}$ with an AIMD algorithm.  A
strawman design would decrease $r_{lim}$ multiplicatively if the link
$L$ is overloaded and stamps the \ldown feedback, or increase it
additively otherwise. However, a malicious sender can manipulate this
design by hiding the \ldown feedback to prevent its rate limit from
decreasing.

To address this problem, we periodically adjust a rate limit, use \lup as a robust signal
to increase the rate limit, and ensure that a sender cannot obtain
valid \lup feedback for a full control interval if its traffic
congests the link $L$. Let $I_{lim}$ denote the control interval
length for rate adjustment on an access router. Suppose a downstream
bottleneck router $R_b$ has a link $L$ in the \bad state.  $R_b$
monitors $L$'s congestion status using a load-based~\cite{vcp-ton} or
a loss-based algorithm such as Random Early Detection
(RED)~\cite{red}. If $R_b$ detects congestion between time $t$ and
$t_1$, it will stamp the \ldown feedback into all packets traversing $L$
from time $t$ until two control intervals after $t_1$: $t_1 + 2
I_{lim}$, even if it has considered the link not congested after
$t_1$. This hysteresis ensures that if a sender congests a link $L$
during one control interval, it will only receive the \ldown feedback in
the following control interval, as shown in
Figure~\ref{fig:robust-signal}.

For each rate limiter $(src, L)$, the access router $R_a$ keeps two
state variables: $t_s$ and $hasIncr$, to track the feedback it
has received. The variable $t_{s}$ records the start time of the rate
limiter's current control interval, and $hasIncr$ records whether the
rate limiter has seen the \lup feedback with a timestamp newer than
$t_{s}$. At the end of each control interval, $R_a$ adjusts the rate
limiter $(src, L)$'s rate limit $r_{lim}$ as follows:

\begin{enumerate*}
\vspace{-0.03in}
\item If $hasIncr$ is true, $R_a$ compares the throughput of the rate
limiter with $\frac{1}{2} r_{lim}$. If the former is larger, $r_{lim}$
will be increased by $\Delta$; otherwise, $r_{lim}$ will remain
unchanged.  Checking the rate limiter's current throughput prevents a
malicious sender from inflating its rate limit by sending slowly for a
long period of time.
\item Otherwise, $R_a$ will decrease $r_{lim}$ to $(1-\delta) r_{lim}$.
\vspace{-0.03in}
\end{enumerate*}
We discuss how to set the parameters $\Delta$, $\delta$, etc. in
\S~\ref{sec:para_settings}.

We now explain why this AIMD algorithm is robust, \ie, a malicious sender
cannot gain unfair bandwidth share by hiding the \ldown feedback: if
a sender has sent a packet when a link $L$ suffers congestion, the
sender's rate limit for $L$ will be decreased. Suppose $L$'s router
$R_b$ detects congestion and starts stamping the \ldown feedback at time
$t$, and let $t_e$ denote the finishing time of an access router's
control interval that includes the time $t$, as shown in
Figure~\ref{fig:robust-signal}.  $R_b$ will stamp the \ldown feedback
between $[t,t_1 + 2 I_{lim}]$. Since $t_e \in [t, t+I_{lim}]$, a
sender will only receive the \ldown feedback for packets sent during
the control interval $[t_e, t_e+I_{lim}]$, because $t_e \geq t$ and
$t_e+I_{lim} < t_1+2I_{lim}$\footnote{This inequation also indicates that
$2I_{lim}$ is the minimal hysteresis to ensure robustness. If the
router $R_b$ stamps the \ldown feedback for shorter than $2I_{lim}$
after $t_1$, an attacker may obtain the \lup feedback in the interval
$[t_e,t_e+I_{lim}]$.}. It can either present the \ldown feedback
newer than $t_e$ to its access router, or present one older than
$t_e$, or not send a packet. All these actions will cause its rate
limit to decrease according to the second rule above.

A legitimate sender should always present \lup feedback to its
access router as long as the feedback has not expired, even if it has
received newer \ldown feedback. This design makes a legitimate
sender mimic an aggressive sender's strategy and ensures fairness
among all senders.

\subsubsection{Handling Multiple Bottlenecks}
\label{sec:multi-ratelimiters}

When a flow traverses multiple links in the \bad state, the flow's
access router will instantiate multiple per-(sender, bottleneck link)
rate limiters for the sender. The present \sys design sends a regular
packet to only one rate limiter for simplicity, but it may overly
limit a sender's sending rate in some cases. This is because when a
sender's packets carry the congestion policing feedback from one of
the bottleneck links, all other rate limiters stay idle. The sender's
access router will reduce their rate limits, if they are idle for
longer than a full control interval, as described above
(\S~\ref{sec:robust-rate-probing}). Consequently, the idle rate
limiters' rate limits may become smaller than a sender's fair share
rates at those bottleneck links. When the bottleneck link carried in a
sender's packets changes, the sender may obtain less than its fair
share bandwidth at the new bottleneck initially, until its rate limit
for the new bottleneck converges. If the bottleneck link in the
packets changes frequently, it is possible that none of the rate
limits converge, giving the sender a throughput smaller than its fair
share bandwidth at any of the bottleneck links. In addition, when the
bottleneck links' rate limits differ greatly and a sender's packets
switch among them frequently, it may be difficult for a transport
protocol such as TCP to adjust a flow's sending rate to match the
abruptly changing rate limit, further reducing a sender's throughput.

We have considered various solutions to address this problem. One
simple solution is to allow a packet to carry all feedback
from all the bottleneck links on its path. An access router can then
pass the packet through all the on-path rate limiters, each receiving
its own feedback and policing the packet independently. This solution
requires a longer and variable-length \sys header. Another one is for
an access router to infer the on-path bottleneck links of a packet
based on history information and send the packet through all the
inferred rate limiters.

We do not include these solutions in the core design for simplicity.
The details of these solutions can be found in
Appendix~\ref{sec:appendix_mbtnk}. Simulation results in
\S~\ref{sec:sim-regular-pkts} and Appendix~\ref{sec:appendix_mbtnk}
suggest that \sys's performance is acceptable. Thus, we consider it a
worthy tradeoff to keep the design simple.

\subsection{Securing Congestion Policing Feedback}
\label{sec:secure_feedback}

Congestion policing feedback must be unforgeable. Malicious end
systems should not be able to forge or tamper the feedback, and
malicious routers should not be able to modify or remove the feedback
stamped by other routers. The \sys design uses efficient symmetric key
cryptography to achieve these goals.

\begin{figure}[t]
\centering
\includegraphics[width=0.9\columnwidth]{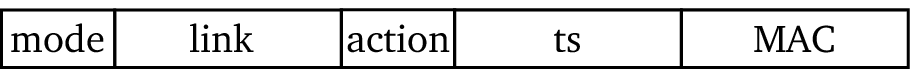}
\caption{\label{fig:feedback}{\bf \small The key 
    congestion policing feedback fields.}}
\end{figure}

\paragraph{Feedback format:} A congestion policing feedback consists of
five key fields as shown in Figure~\ref{fig:feedback}: $mode$, $link$,
$action$, $ts$, and $MAC$. When the $mode$ field is \ok, it represents
the \ok\ feedback. When the $mode$ field is \bad, the $link$ field
indicates the identifier (an IP address) of the corresponding link
$L$, and the $action$ field indicates the detailed feedback: if
$action$ is \lo (\hi), it is the \lup (\ldown) feedback.  The $ts$
field records a timestamp, and the $MAC$ field holds a MAC signature
that attests the feedback's integrity.

In addition to the five key fields, a \bad feedback also includes a
field $token_{nop}$. We explain the use of this field later in this
section.

\paragraph{Stamping \ok\ feedback:} When an access router stamps the
\ok\ feedback, it sets $mode$ to \ok, $link$ to a null identifier
$link_{null}$, $action$ to \lo, $ts$ to its local time, and uses a
time-varying secret key $K_a$ known only to itself to compute the
$MAC$:
\begin{equation}
token_{nop} = MAC_{K_a}(src, dst, ts, link_{null}, nop) \label{eq:nop}
\end{equation}

The MAC computation covers both the source and destination addresses
to prevent an attacker from re-using valid \ok feedback on a different
connection.

\paragraph{Stamping \lup\ feedback:} When an access router stamps the
\lup\ feedback, the $mode$ field is already \bad, and the $link$ field
already contains the link identifier $L$. The router sets $action$ to
\lo\ and $ts$ to its local time, and computes the $MAC$ field using
the secret key $K_a$:
\begin{equation}
token_{L^\uparrow} = MAC_{K_a}(src, dst, ts, L, mon, incr) \label{eq:incr}
\end{equation}

The router also inserts a $token_{nop}$ as computed in Eq~(\ref{eq:nop})
into the $token_{nop}$ field.

\paragraph{Stamping \ldown feedback:} When a link $L$'s router
$R_b$ stamps the \ldown feedback, it sets $mode$ to \bad, $link$ to
$L$, $action$ to \hi, and computes a new $MAC$ value using a secret
key $K_{ai}$ shared between its AS and the sender's AS:
\begin{equation}
\small token_{L^\downarrow} = MAC_{K_{ai}}(src, dst, ts, L, mon, decr,
token_{nop}) \label{eq:decr}
\end{equation}

The shared secret key $K_{ai}$ is established by piggybacking a
distributed Diffie-Hellman key exchange in BGP as
in~\cite{passport}. The router $R_b$ includes $token_{nop}$ stamped by
the sender's access router in its MAC computation, and erases it
afterwards to prevent malicious downstream routers from overwriting
its feedback.

If $R_b$ is an AS internal router that does not speak BGP, it may not
know $K_{ai}$. In this case, $R_b$ can leave the $MAC$ and
$token_{nop}$ fields unmodified and let an egress border router of the
AS update their values when the packet exits the AS. This design
reduces the management overhead to distribute $K_{ai}$ to an internal
router $R_b$.

\paragraph{Validating feedback:} When a source access router receives a
regular packet, it first validates the packet's congestion policing
feedback. If the feedback is invalid, the packet will be treated as a
request packet and subject to per-sender request packet policing.

A feedback is considered invalid if its $ts$ field is more than $w$
seconds older than the access router's local time $t_{now}$:
$|t_{now}-ts|>w$, or if the $MAC$ field has an invalid signature. The
$MAC$ field is validated using Eq~(\ref{eq:nop}) and
Eq~(\ref{eq:incr}) for the \ok and \lup feedback, respectively. To
validate \ldown feedback, the access router first re-computes the
$token_{nop}$ using Eq~(\ref{eq:nop}), and then re-computes the MAC
using Eq~(\ref{eq:decr}).  The second step requires the access router
to identify the link $L$'s AS in order to determine the shared secret
key $K_{ai}$. We can use an IP-to-AS mapping
tool~\cite{Mao2003} for this purpose, as the feedback
includes the link $L$'s IP address.

\subsection{Localizing Damage of Compromised Routers}
\label{sec:localize_damage}

The \sys\ design places enforcement functions that include feedback
validation and traffic policing at the edge of the network to be
scalable. However, if an access router is compromised, attackers in
its subnet or itself may misbehave to congest the network. \sys
addresses this problem by localizing the damage to the compromised
AS. If an AS has a compromised router, we consider the AS as
compromised, and do not aim to provide guaranteed network access for
that AS's legitimate traffic.

A \sys router can take several approaches to localize the damage of
compromised ASes, if its congestion persists after it has started a
monitoring cycle, a signal of malfunctioning access routers. One
approach is to separate each source AS's traffic into different queues. This
requires per-AS queuing.  We think the overhead is affordable because
the present Internet has only about 35K ASes~\cite{bgp-report}. We may
replace per-AS queuing with per-AS rate limiting and set the rate
limits by periodically computing each AS's max-min fair share
bandwidth on the congested link as in~\cite{pushback}. Another more
scalable approach is to use a heavy-hitter detection algorithm such as
RED-PD~\cite{red-pd} to detect and throttle high-rate source ASes. A
heavy-hitter detection algorithm is suitable in this case because
legitimate source ASes will continuously reduce their senders' traffic
as long as they receive the \ldown feedback. The detected high-rate
ASes are likely to be the compromised ASes that do not slow down their
senders.

All these approaches require a router to correctly identify the source
AS of a packet, which can be achieved using an IP-to-AS mapping tool if the
source IP address of the packet is not spoofed. \sys uses
Passport~\cite{passport} to prevent source address spoofing. A
Passport header is inserted between IP and the \sys header.  Passport
piggybacks a distributed Diffie-Hellman key exchange in the
inter-domain routing system to enable each pair of ASes to share a
secret key. A source AS uses a key it shares with an AS on the path to
a destination to compute a secure MAC and inserts it into a packet's
Passport header. Each AS on the path can verify that a packet is
originated from the source AS by validating the corresponding MAC.
\sys also uses Passport to establish the shared secret keys between
ASes to secure the congestion policing feedback
(\S~\ref{sec:secure_feedback}).

\subsection{Parameter Settings}
\label{sec:para_settings}

Figure~\ref{fig:parameter} summarizes the main parameters in the
\sys\ design and their values used in our implementation.  The level-1
request packets ($l_1$) are rate limited at one per 1 ms. A request
packet size is estimated as 92 bytes that includes a 40-byte TCP/IP
header, a 28-byte \sys header (Figure~\ref{fig:pkthdr}) and a 24-byte
Passport header~\cite{passport}. We set the
control interval $I_{lim}$ to 2 seconds, one order of magnitude larger
than a typical RTT ($<$ 200ms) on the Internet. This allows an
end-to-end congestion control mechanism such as TCP to reach a
sender's rate limit during one control interval. We do not further
increase $I_{lim}$ because a large control interval would slow down
the rate limit convergence.

The rate limit AI parameter $\Delta$ can be neither too small nor too
large: a small $\Delta$ would lead to slow convergence to fairness; a
large $\Delta$ may result in significant overshoot. We set $\Delta$ to
12Kbps because it works well for our targeted fair share rate range:
50Kbps $\sim$ 400Kbps. A legitimate sender may abort a connection if
its sending rate is much lower than 50Kbps, and 400Kbps should
provide reasonable performance for a legitimate sender during DoS
flooding attacks.  The rate limit MD parameter $\delta$ is set to 0.1,
a value much smaller than TCP's MD parameter 0.5.  This is because a
router may stamp the \ldown feedback for two control intervals longer
than the congestion period (\S~\ref{sec:robust-rate-probing}).

We set the attack detection threshold $p_{th}$ to 2\%, since at this packet loss
rate, a TCP flow with 200ms RTT and 1500B packets can obtain
about 500Kbps throughput~\cite{tcp-throughput-equation}. We set a
link's maximum queue length $Q_{lim}$ to 200ms $\times$ the link's
capability. We use a loss-based algorithm RED to detect a link's
congestion status. It is our future work to implement a
load-based algorithm (\eg, \cite{vcp-ton}).

\section{Analysis}
\label{sec:analysis}

In this section, we analyze the scalability and security of \sys, and
discuss the incremental deployment issues.

\subsection{Scalability}
\label{sec:scalability}

As a closed-loop design, \sys can place different functions at
different locations to provide per-sender fairness. It places
per-sender traffic policing at access routers, and lightweight
congestion detection, feedback stamping, and AS-level policing at
bottleneck routers. In contrast, per-host fair queuing, an open-loop
solution used in previous work~\cite{tva-ton,stopit}, does not have
this flexibility. Every bottleneck router must keep per-host queues to
provide per-sender (or per-receiver) fairness. There are only 35K ASes
on today's Internet~\cite{bgp-report}, while the number of compromised
hosts involved in a DoS attack could reach
millions~\cite{conficker-worm-9m}.  Thus, compared to per-host fair
queuing, \sys can significantly reduce the amount of state kept by a
bottleneck router.

However, \sys access routers need to perform per-(sender, bottleneck
link) rate limiting. Our calculation suggests that with today's hardware
technology, they can afford to do so and will not
become a new scaling bottleneck. While we do not have accurate data to
estimate the number of bottlenecks a sender traverses during attack
times, we think 100 links per legitimate sender is a reasonable upper
bound. An access router can aggregate a sender's rate limiters by
bottleneck links' prefixes if a sender needs more than 100 rate
limiters. If an access router serves 10K end hosts, it will have
at most one million rate limiters in total. Each rate
limiter needs about 24 bytes of memory for state variables
(1 bit for $hasIncr$, 8 bytes for two timestamps, 4 bytes for the rate limit, and 12 bytes for a queue
object) and another 1500 bytes to queue at least one packet.  The
total amount of memory requirement is less than 2GB, and we can use
fast DRAM for this purpose as access routers' line speeds are
typically slower than those of core routers.

The processing overhead of an access router is also acceptable.  The
per-packet processing time on our benchmarking PC is less than $1.3
\mu s$ during attack times (\S~\ref{sec:benchmarking}).  This
translates into a throughput of 770K packets per second, or more than
9\,Gbps, assuming 1500-byte packet size and CPU is the throughput
bottleneck. Implementing the cryptographic operations in hardware can
further improve an access router's throughput.

\subsection{Security}
\label{sec:security}

Next we summarize how \sys withstands various attacks.

\subsubsection{Malicious End Systems}

\paragraph{Forgery or Tampering:} Malicious end systems  may
attempt to forge valid congestion policing feedback. But \sys
protects congestion policing feedback with MAC signatures. As long
as the underlying MAC is secure, malicious end systems cannot spoof
valid feedback. A malicious sender may selectively present \lup or
hide \ldown to its access router, but \sys's robust AIMD
algorithm (\S~\ref{sec:robust-rate-probing}) prevents it from gaining
a higher rate limit.

\paragraph{Evading attack detection:} Malicious end systems may attempt to
prevent a congested router from starting a monitoring cycle. This
attack is ineffective when a well-provisioned router uses high link
utilization to detect attacks. When an under-provisioned router uses
the packet loss rate to detect attacks, \sys limits the damage of this
attack with a low loss detection threshold $p_{th}$
(\S~\ref{sec:monitoring}).

\paragraph{On-off attacks:} Attackers may attempt to launch
on-off attacks. In a macroscopic on-off attack, attackers may flood
the network again after a congested router terminates a monitoring
cycle. \sys uses a prolonged monitor cycle (\S~\ref{sec:monitoring})
to mitigate this attack. In a microscopic on-off attack, attackers may
send traffic bursts with a short on-off cycle, attempting to congest
the network with synchronized bursts, yet maintaining average sending
rates lower than their rate limits. Our theoretical bound in
\S~\ref{sec:arch} and simulation results in
\S~\ref{sec:sim-regular-pkts} both show that the shape of attack
traffic cannot reduce a legitimate user's guaranteed bandwidth share,
because a sender cannot send faster than its rate limit at any time
(\S~\ref{sec:police-regular}), and \sys's robust rate limit adjustment
algorithm (\S~\ref{sec:robust-rate-probing}) prevents a sender from
suddenly increasing its actual sending rate.

\subsubsection{Malicious On-path Routers}

A malicious router downstream to a congested link may attempt to
remove or modify the \ldown feedback stamped by a congested router
in order to hide upstream congestion. But such attempts will make the
feedback invalid, because the router does not know the original
$token_{nop}$ value needed to compute a valid MAC
(\S~\ref{sec:secure_feedback}).

A malicious on-path router may discard packets to completely disrupt
end-to-end communications, duplicate packets, or increase packet sizes
to congest downstream links. It may also change the request packet
priority field in a \sys header to congest the request channel on
downstream links. Preventing such attacks requires Byzantine tolerant
routing~\cite{perlman-phd}, which is not \sys's design goal. Instead,
we aim to make these attacks detectable. Passport~\cite{passport}, the
source authentication system \sys uses, partially protects the
integrity of a packet and enables duplicate detection. It includes a
packet's length and the first 8 bytes of a packet's transport payload
(which includes the TCP/UDP checksum) in its MAC computation. We can
further extend Passport's MAC computation to include \sys's request
packet priority field to protect it.

\subsection{Incremental Deployment}
\label{sec:deployment}

\sys can be incrementally deployed by end systems and routers. Since
the \sys header is a shim layer between IP and upper layer protocols,
legacy applications need not be modified. Legacy routers can ignore
the \sys header and forward packets using the IP header.  Routers at
congested links and access routers need to be upgraded, but
well-provisioned routers that can withstand tens of Gbps attack
traffic may not need to upgrade. The deployment can take a
bump-in-the-wire approach, by placing inline boxes that implement
\sys's enforcement functions in front of the routers that require
upgrading.  Middleboxes such as firewalls need to be configured to
permit \sys traffic.

\sys provides deployment incentives to both end systems and ASes,
because legacy traffic is treated by deployed ASes with lower priority
( Figure~\ref{fig:queueing}). Deployed ASes can form a trusted overlay
network and protect each other's legitimate traffic within their
networks. Their traffic is not protected at undeployed networks,
encouraging them to direct traffic to other deployed ASes using BGP.

\section{Implementation and Evaluation}
\label{sec:eval}

\begin{figure}[t]
\centering
\includegraphics[width=0.95\columnwidth]{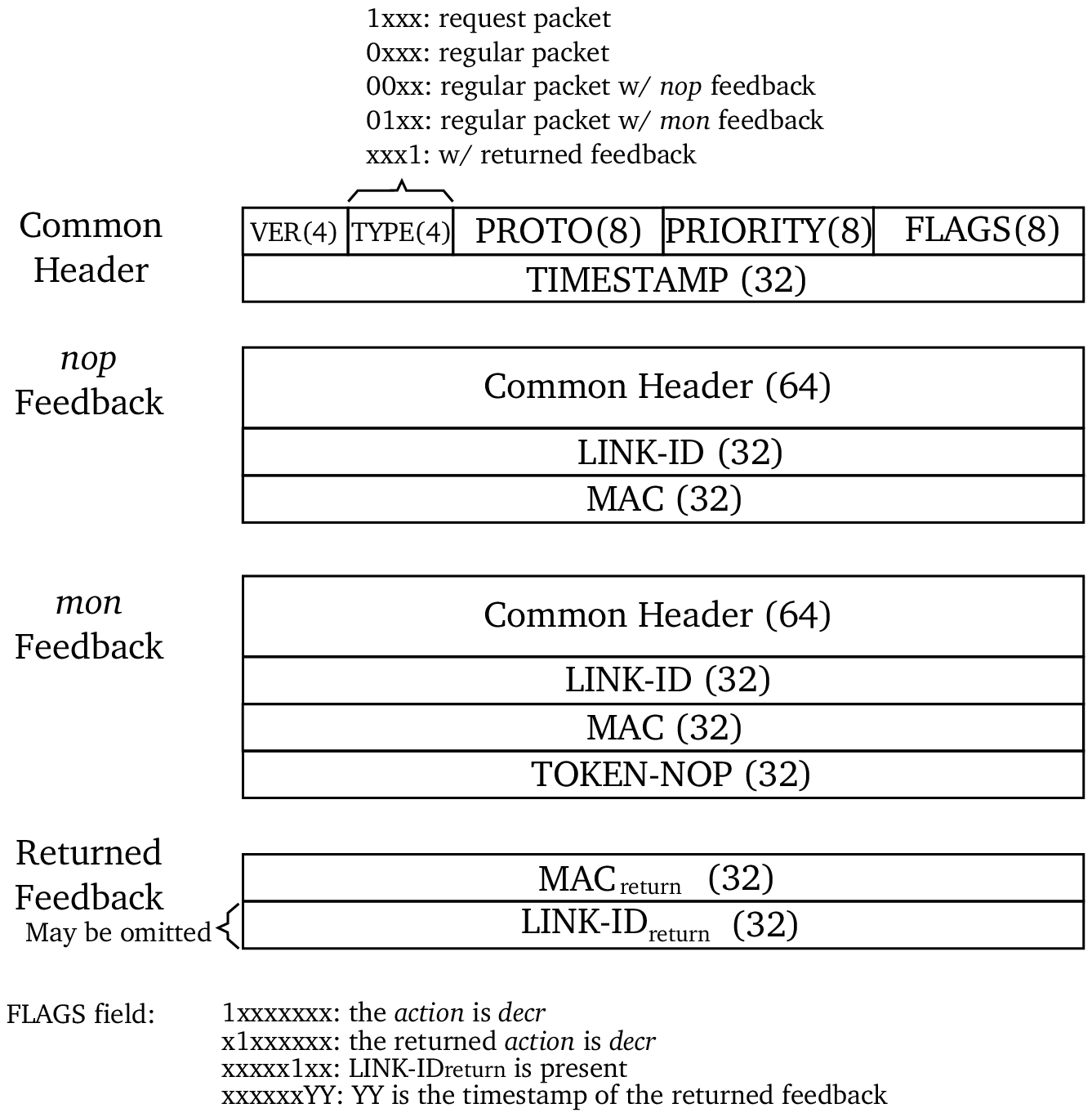}
\caption{\label{fig:pkthdr} {\bf \small The \sys\ header format.}}
\end{figure}

We have implemented \sys prototypes in Linux and in the ns-2
simulator. Next we evaluate the \sys header and packet processing
overhead with our Linux implementation, and use ns-2 simulations to
show how effective \sys mitigates DoS attacks.

\subsection{\sys Header}
\label{sec:pkthdr}

Figure~\ref{fig:pkthdr} shows the format of a \sys header in our Linux
implementation. A full \sys header from a sender to a receiver
includes a forward header and a return header. The forward header
includes the congestion policing feedback on the forward path from the
sender to the receiver, and the return header includes the reverse
path information from the receiver to the sender. Most fields are
self-explained. A \sys header is implemented as a shim layer between
IP and an upper-layer protocol, and the PROTO field describes the
upper-layer protocol (\eg, TCP or UDP). The unit of a timestamp is one
second.

The return header may be omitted to reduce overhead if the sender has
previously returned the latest feedback to the receiver. Even when the
return header is present, it does not always include all the fields.
If the returned feedback is \ok, the LINK-ID$_{return}$ field will be
omitted because it is zero, and one bit in the FLAGS field indicates
this omission.

A \sys header only includes the last two bits of the returned
timestamp to save header space. In the subsequent packets from the
sender to the receiver, the sender's access router will reconstruct
the full timestamp from its local time and the returned two bits,
assuming that the timestamp is less than four seconds older than its
current time. With this implementation, a \sys header is 20 bytes in the common case
when the feedback is \ok for both the forward and return paths. In
the worst case that the feedback is \bad for both paths, the header is
28 bytes long.

\subsection{Micro-benchmarking}
\label{sec:benchmarking}

We have implemented \sys in Linux using XORP~\cite{handley05:xorp} and
Click~\cite{click}. We modified XORP's BGP module to establish the
pairwise symmetric keys shared between ASes. We added the data packet
processing logic into Click and ran Click routers in the kernel space
for packet forwarding. XORP communicates with Click via the
\texttt{/click} file system.  We added a module between the IP and
transport layers on end-hosts to handle \sys headers. This design
keeps upper-layer TCP/UDP protocols and legacy applications
unmodified. We use AES-128 as a secure MAC function due to its fast
speed and available hardware support~\cite{aes-hardware,intel-aes}.

We benchmark the Linux implementation on Deterlab~\cite{deterlab} with
a three-node testbed. A source access router $A$ and a destination $C$
are connected via a router $B$. The $B$---$C$ link is the bottleneck
with a capacity of 5Mbps. Each node has two Intel Xeon 3GHz CPUs and
2GB memory. To benchmark the processing overhead without attacks,
we send 100Kbps UDP request packets and 1Mbps UDP regular
packets from $A$ to $C$ respectively. To benchmark the overhead in
face of DoS attacks, we send 1Mbps UDP request packets and
10Mbps UDP regular packets simultaneously.

\begin{figure}[t]
\centering
\small
\begin{tabular}{l|l|l|l}
Packet & Router & \multicolumn{2}{c}{Processing Overhead (ns/pkt)} \\ \cline{3-4}
Type & Type & \sys & TVA+ \\ \hline
\multirow{3}{*}{request} & \multirow{2}{*}{bottleneck} & w/o attack: 0 & \multirow{3}{*}{389} \\
& & w/ attack: 492 & \\ \cline{2-3}
& access & 546 & \\ \hline
\multirow{4}{*}{regular} & \multirow{2}{*}{bottleneck} & w/o attack: 0 & \\
& & w/ attack: 554 & 
\\ \cline{2-3}
& \multirow{2}{*}{access} & w/o attack: 781 & 791 \\
& & w/ attack: 1267 &
\end{tabular}
\vspace{-1ex}
\caption{\label{fig:bench} {\bf \small \sys implementation
    micro-benchmarking results.}}
\vspace{-0.1in}
\end{figure}

The benchmarking results are shown in Figure~\ref{fig:bench}. With
\sys, when there is no attack, a request packet does not need any
extra processing on the bottleneck router $B$, but it introduces an
average overhead of 546ns on the access router $A$ because the router
must stamp the \ok feedback into the packet. A regular packet does not
incur any extra processing overhead on the bottleneck router either,
but it takes the access router 781ns on average to validate the returned
feedback and generate a new one. When the bottleneck link enters the
\bad state during attack times, the bottleneck router takes 492ns to
process a 92B request packet, or at most 554ns to process a 1500B
regular packet. The access router takes on average 1267ns to process a
regular packet at attack times.

The performance of a capability system TVA+~\cite{stopit} on the same
topology is also shown in Figure~\ref{fig:bench} for comparison. We
can see that the processing overhead introduced by \sys is on par with
that of TVA+. Note that we do not show the result when TVA+ caches
capabilities, because such caching requires per-flow state on routers,
while \sys does not have this requirement.

These results show that \sys's per-packet overhead is low. The
CPU-intensive operations are primarily AES computation. Since there
exists commercial hardware that can support AES operations at
40Gbps~\cite{aes-hardware}, we expect that \sys's per-packet
processing will not become a performance bottleneck. We note that the
benchmarking results do not include the Passport
overhead, as a Passport header can be updated by
inline boxes near an AS's ingress and egress border routers~\cite{passport}.

\subsection{Mitigating DoS Flooding Attacks}
\label{sec:simulations}

Next we evaluate how well \sys mitigates various DoS flooding attacks using
ns-2 simulations. We also compare \sys with three other representative
DoS mitigation schemes:

\paragraph{TVA+}: TVA+~\cite{tva-ton,stopit} is a  network
architecture that uses network capabilities and per-host fair queuing
to defend against DoS flooding attacks. TVA+ uses two-level
hierarchical fair queuing
(first based on the source AS and then based on the source IP address)
at congested links to mitigate request packet flooding attacks, and
per-receiver fair queuing to mitigate authorized traffic flooding
attacks in case (colluding or incompetent) receivers fail to stop
attack traffic.

\paragraph{StopIt}: StopIt~\cite{stopit} is a filter and fair queuing
based DoS defense system.  A targeted victim can install network filters to
stop unwanted traffic. Similar to TVA+, in case
receivers fail to install filters, StopIt uses hierarchical queuing
(first based on the source AS and then based on the source IP address)
at congested links to separate legitimate traffic from attack traffic.

\paragraph{Fair Queuing (FQ)}: Per-sender fair queuing at every link
provides a sender its fair share of the link's bandwidth. We use fair
queuing to represent a DoS defense mechanism that aims to throttle
attack traffic to consume no more than its fair share of bandwidth.

We have implemented TVA+ and StopIt as described in~\cite{stopit,tva-ton}. We
use the Deficit Round Robin (DRR) algorithm~\cite{deficit-round-robin}
to implement fair queuing because it has $O(1)$ per packet operation
overhead. In our simulations, attackers do not spoof source addresses
because \sys uses Passport~\cite{passport} to prevent spoofing. Thus,
routers could queue attack traffic separately from legitimate traffic.

\subsubsection{Unwanted Traffic Flooding Attacks}
\label{sec:sim-init-pkts}

\begin{figure}[t]
\centering
\includegraphics[width=0.9\columnwidth]{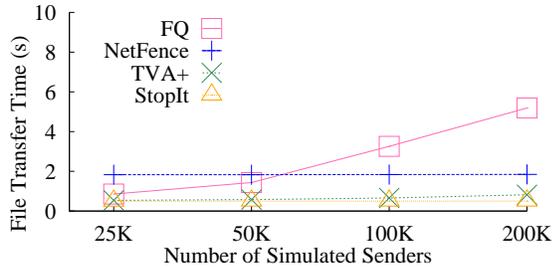}
\caption{\label{fig:init-flooding} {\bf \small The average transfer
    time of a 20KB file when the targeted victim can identify and wish
    to remove the
    attack traffic. The file transfer completion ratio is 100\% in
    all simulated systems.}}
\end{figure}

We first simulate the attack scenario where the attackers directly
flood a victim, but the victim can classify the attack traffic, and
uses the provided DoS defense mechanism: capabilities in TVA+, secure
congestion policing feedback in \sys, and filters in StopIt, to block the
unwanted traffic.

We desire to simulate attacks in which thousands to millions of
attackers flood a well provisioned link. However, we are currently
unable to scale our simulations to support beyond several thousand
nodes. To address this limitation, we adopt the evaluation approach
in~\cite{tva-ton}. We fix the number of nodes, but scale down the
bottleneck link capacity proportionally to simulate the case where the
bottleneck link capacity is fixed, but the number of attackers
increases.

We use a dumb-bell topology in which ten source ASes connect to a
destination AS via a transit AS. Each source AS has 100 source hosts
connected to a single access router. The transit AS has two routers
$R_{bl}$ and $R_{br}$, and the destination AS has one victim
destination host. The link between $R_{bl}$ and $R_{br}$ is the
bottleneck link, and all other links have sufficient capacity to
avoid congestion. We vary the bottleneck link capacity from 400Mbps to
50Mbps to simulate the scenario where 25K $\sim$ 200K senders (both
legitimate and malicious) share a 10Gbps link. Each sender's fair
share bandwidth varies from 400Kbps $\sim$ 50Kbps, which is \sys's
targeted operating region. The propagation delay of each link is 10ms.

In the simulations, each sender is either a legitimate user or an
attacker. To stress-test our design, we let each source AS have only
one legitimate user that repeatedly sends a 20KB file to the victim
using TCP. We let each attacker send 1Mbps constant-rate UDP traffic
to the victim. We measure the effectiveness of a DoS defense system
using two metrics: 1) the average time it takes to complete a
successful file transfer; and 2) the fraction of successful file
transfers among the total number of file transfers initiated. We set
the initial TCP SYN retransmission timeout to 1 second, and abort a
file transfer if the TCP three-way handshake cannot finish after nine
retransmissions, or if the entire file transfer cannot finish in 200
seconds. We terminate a simulation run when the simulated time reaches
4000 seconds.

For each DoS defense system we study, we simulate the most effective
DoS flooding attacks malicious nodes can launch. In case of an
unwanted traffic flooding attack, the most effective flooding strategy
in \sys and TVA+ is the request packet flooding attack.  Under this
attack, each \sys sender needs to choose a proper priority level for
its request packets. We make an attacker always select the highest
priority level at which the aggregate attack traffic can saturate the
request channel. A legitimate sender starts with the lowest priority
level and gradually increases the priority level if it cannot obtain
valid congestion policing feedback.

Figure~\ref{fig:init-flooding} shows the simulation results. The
average file transfer completion ratio is omitted because all file
transfers complete in these simulations. As can be seen, StopIt has
the best performance, because the attack traffic is blocked near the
attack sources by network filters. TVA+ and \sys also have a short
average file transfer time that only increases slightly as the number
of simulated senders increases. This is because in a request packet
flooding attack, as long as a legitimate sender has one request packet
delivered to the victim, it can send the rest of the file using
regular packets that are not affected by the attack traffic. The
average file transfer time in \sys is about one second longer than
that in TVA+, because a legitimate sender will initially send a
level-0 request packet that cannot pass the bottleneck link due to
attackers' request packet floods. After one second retransmission
backoff, a sender is able to retransmit a request packet with
sufficiently high priority (level-10) to pass the bottleneck link.
Attackers cannot further delay legitimate request packets, because
they are not numerous enough to congest the request channel at this
priority level.

Figure~\ref{fig:init-flooding} also shows that FQ alone is an
ineffective DoS defense mechanism. With FQ, the average file transfer
time increases linearly with the number of simulated senders, as each
packet must compete with the attack traffic for the bottleneck
bandwidth.

These results show that \sys performs similarly to capability-based
and filter-based systems when targeted victims can stop the attack
traffic.  A legitimate sender may wait longer in \sys to successfully
transmit a request packet than in TVA+ or StopIt. This is because \sys
uses coarse-grained exponential backoff to schedule a request packet's
transmission and set its priority, while TVA+ uses fine-grained but
less scalable per-sender fair queuing to schedule a request packet's
transmission, and StopIt enables a victim to completely block unwanted traffic.

\subsubsection{Colluding Attacks}
\label{sec:sim-regular-pkts}

Next we present our simulation results for regular traffic flooding
attacks where malicious sender-receiver pairs collude to flood the
network. Such attacks may also occur if DoS victims fail to identify
the attack traffic.

\begin{figure}[t!]
\subfigure[Long-running TCP]{\label{fig:data-flooding-long-75}
\begin{minipage}{\columnwidth}
\centering
\includegraphics[width=0.9\columnwidth]{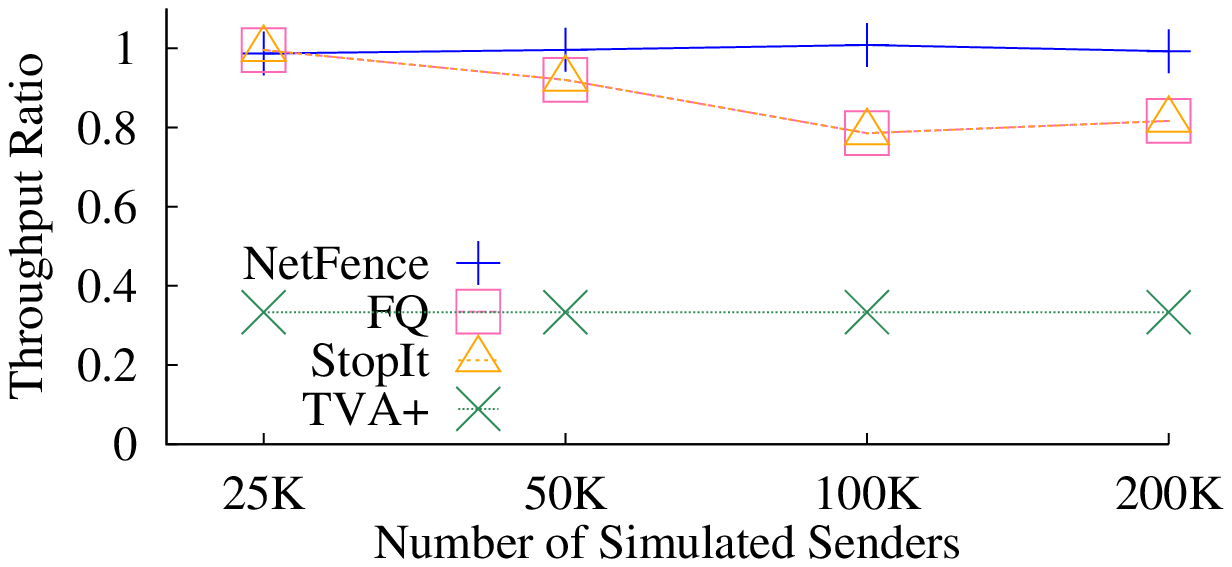}
\end{minipage}}
\subfigure[Web-like traffic]{\label{fig:data-flooding-short-75}
\begin{minipage}{\columnwidth}
\centering
\includegraphics[width=0.9\columnwidth]{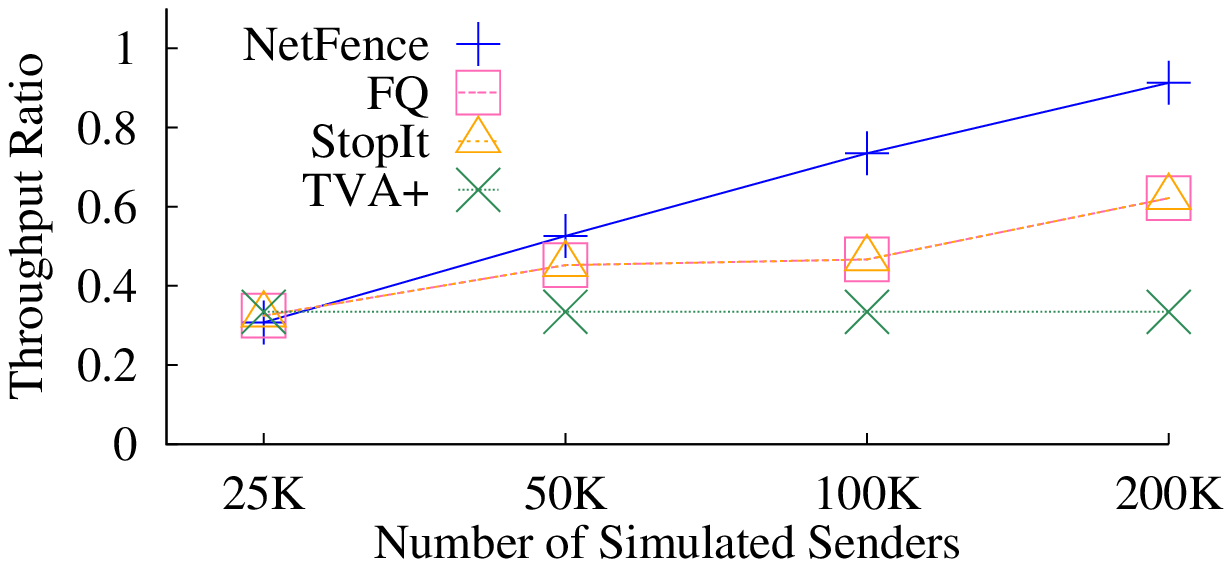}
\end{minipage}}
\caption{\label{fig:data-flooding-75} {\bf \small {\em Throughput
      Ratio} between legitimate users and attackers when
    receivers fail to suppress the attack traffic. {\em Fairness
      Index} among legitimate users is close to 1 in all the
    simulations.}}
\end{figure}

\paragraph{Single Bottleneck:} We use a similar topology as in
the previous experiments (\S~\ref{sec:sim-init-pkts}) to simulate
colluding attacks. In this simulation topology, the router at the
right-hand side of the bottleneck link $R_{br}$ connects to one
destination AS with a victim host and nine additional ASes, each
having a colluding host (colluder). Each source AS has 25\% legitimate
users and 75\% attackers, simulating the case where the attackers are
numerous but there are still a reasonable number of legitimate users
in each source AS.

Each legitimate user sends TCP traffic to the victim host. We simulate
two types of user
traffic: 1) long-running TCP, where a legitimate sender sends a single
large file; 2) web-like traffic, where a sender
sends small files whose size distribution mimics that of web
traffic. We draw the file size distribution from a mixture of Pareto
and exponential distributions as in~\cite{luo05}, and make the
interval between two file transfers uniformly distributed between 0.1
and 0.2 seconds. The maximum file size is limited to 150KB to
make the experiments finish within a reasonable amount of time.

To simulate colluding attacks, we let each attacker send 1Mbps UDP
traffic to a colluder.  The attackers in TVA+ and \sys send regular
packets. Colluders in StopIt do not install filters to stop the attack
traffic. We simulate each experiment for 4000 seconds.

When compromised nodes organize into pairs to send attack traffic, \sys aims to
guarantee each legitimate sender its fair share of the bottleneck
bandwidth without keeping per-sender queues in the core network. We
use two metrics to measure a DoS defense system's performance under
this type of attack: 1) \textit{Throughput Ratio}, the ratio between
the average throughput of a legitimate user and that of an attacker;
and 2) \textit{Fairness Index} among legitimate users~\cite{aimd}. Let
$x_i$ denote a legitimate sender $i$'s throughput, and the fairness
index is defined as $(\sum x_i)^2/(n \sum x^2_i)$.  The ideal
throughput ratio is 1, indicating that a legitimate user obtains on
average the same bottleneck bandwidth as an attacker. The ideal
fairness index is also 1, indicating that each legitimate sender has
the same average throughput. We only measure the fairness index among
legitimate users because \textit{Throughput Ratio} has already
quantified how well a legitimate user performs relatively to an
attacker.

Figure~\ref{fig:data-flooding-75} shows the simulation results. The
fairness index for all systems is close to 1 in all the simulations
and is thus not shown in the figure. For long-running TCP, \sys's
throughput ratio is also close to 1. This result shows that \sys
provides a legitimate sender its fair share of bandwidth despite the
presence of DoS flooding traffic, consistent with the theoretic
analysis in \S~\ref{sec:arch_theorem}. For the web-like traffic,
\sys's throughput ratio increases gradually from 0.3 to close to 1 as
the number of simulated senders increases. The throughput ratio is low
when the number of senders is small, because a legitimate sender
cannot fully utilize its fair share bandwidth: each sender has a large
fair share of bandwidth, but a legitimate sender's web-like traffic
has insufficient demand and there are gaps between consecutive file
transfers.

FQ and StopIt perform exactly the same, because in these colluding
attacks, they both resort to per-sender fair queuing to protect a
legitimate user's traffic. However, unexpectedly, we note that they
provide legitimate users less throughput than attackers even when the
user traffic is long-running TCP. By analyzing packet traces, we
discover that this unfairness is due to the interaction between TCP
and the DRR algorithm.  A TCP sender's queue does not always have
packets due to TCP's burstiness, but a constant-rate UDP sender's
queue is always full. When a TCP sender's queue is not empty, it
shares the bottleneck bandwidth fairly with other attackers, but when
its queue is empty, the attack traffic will use up its bandwidth
share, leading to a lower throughput for a TCP sender.

TVA+ has the lowest throughput ratio among all systems in this
simulation setting, indicating that a small number of colluders can
significantly impact TVA+'s performance. This is because TVA+ uses
per-destination fair queuing on the regular packet channel. With $N_C$
colluders, a DoS victim obtains only $\frac{1}{N_C+1}$ fraction
of the bottleneck capacity $C$ at attack times, and each of the
victim's $G$ legitimate senders obtains $\frac{1}{G(1+N_C)}$ fraction
of the capacity $C$. The attackers, however, obtain an aggregate
$\frac{N_C}{(1+N_c)}$ fraction of $C$. If this bandwidth is shared by
$B$ attackers fairly, each will get a $\frac{N_C}{B(1+N_c)}$ fraction
of the bottleneck capacity.  A sender's bottleneck bandwidth share in
other systems (\sys, StopIt, and FQ) is $\frac{1}{G+B}$, and does not
depend on the number of colluders $N_C$. In our simulations, $N_c =
9$, $G = 25 \% \times 1000$, and $B= 75\% \times 1000$. A
legitimate TVA+ sender obtains $\frac{1}{2500}$ of the bottleneck
bandwidth, while an attacker obtains $\frac{9}{7500}$ of the
bottleneck bandwidth, three times higher than a legitimate sender, as
shown in Figure~\ref{fig:data-flooding-75}.

In these simulations, we also measure the bottleneck link
utilization. The result shows that the utilization is above 90\% for
\sys, and almost 100\% for other systems. \sys does not achieve full
link utilization mainly because a router stamps the \ldown\ feedback for
two extra control intervals after congestion has abated, as explained
in \S~\ref{sec:robust-rate-probing}.

\begin{figure}[t]
\centering
\includegraphics[width=0.9\columnwidth]{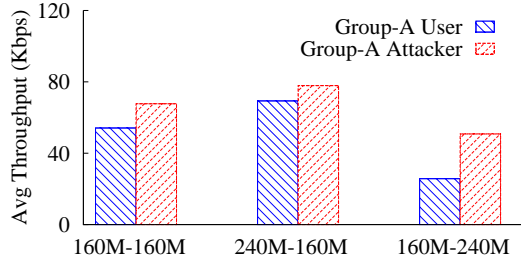}
\caption{\label{fig:multi-data-flooding} {\bf \small Sender throughput
(Kbps) under regular packet flooding attacks in a parking-lot topology
with two bottleneck links. The fair share rate for each sender is
80Kbps.}}
\end{figure}

\paragraph{Multiple Bottlenecks:} To evaluate \sys's performance with
multiple bottlenecks, we simulate colluding attacks on a parking-lot
topology that has two bottleneck links in the \bad\ state: $L_1$ and
$L_2$. 3000 senders are organized into three groups of the same size:
Group-A senders send through both $L_1$ and $L_2$, Group-B senders
send through the second link $L_2$, and Group-C senders send through
the first link $L_1$. Each group contains 75\% attackers and 25\%
legitimate users. Each attacker sends UDP traffic in regular packets
at 1Mbps to a colluder, while each user sends long-running TCP traffic
to a victim. Every simulation terminates at 4000 seconds in simulated
time.

We simulate three different pairs of bottleneck capacities: 1)
$C_{L_1}=C_{L_2}=160Mbps$; 2) $C_{L_1}=240Mbps, C_{L_2}=160Mbps$; and
3) $C_{L_1}=160Mbps, C_{L_2}=240Mbps$. In these cases, a Group-A
sender's max-min fair share bandwidth is 80Kbps, while a Group-B or
Group-C sender's max-min fair share is either 80Kbps or 160Kbps. The
simulation results are shown in Figure~\ref{fig:multi-data-flooding}.
The x-axis shows different simulation cases, and the y-axis is a
sender's average throughput. Senders in Group-B and Group-C are
omitted because they each can get at least its fair share bandwidth in
all the simulations. However, on average a sender in Group-A obtains
much smaller throughput than its fair share rate when
$C_{L_1}<C_{L_2}$. This is because traffic from a Group-A sender
traverses both bottleneck links $L_1$ and $L_2$. As discussed in
\S~\ref{sec:multi-ratelimiters}, if a flow traversing both $L_1$ and
$L_2$ switches between the two corresponding rate limiters frequently,
its rate limit may become smaller than its fair share bandwidth. This
problem does not significantly affect the first two cases in which
$C_{L_1} \ge C_{L_2}$, because in these simulations a Group-A sender's
traffic carries $L_1$'s feedback most of the time. As a result, the
rate limiter $(src,L_1)$ for a Group-A sender $src$ is not idle most
of the time, and it can have a rate limit close to the fair share
bandwidth of the sender $src$.

Figure~\ref{fig:multi-data-flooding} also shows that a TCP user in
Group-A has an even lower average throughput than an attacker in
Group-A (the rightmost case). This is because when a TCP flow switches
between two rate limiters that have very different rate limits, TCP
cannot catch up with the abrupt rate limit change.

\sys's performance with multiple bottleneck links can be improved with
more complicated designs, as discussed in
\S~\ref{sec:multi-ratelimiters} and Appendix~\ref{sec:appendix_mbtnk}.
We believe the current design is a reasonable tradeoff to balance
between complexity and performance.

\begin{figure}[t]
\centering
\includegraphics[width=0.9\columnwidth]{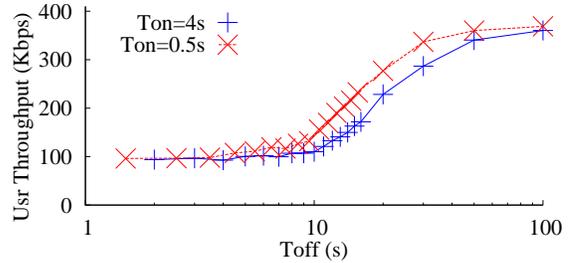}
\caption{\label{fig:sim-on-off} {\bf \small Average user throughput in
face of microscopic on-off attacks. The user traffic is long-running
TCP. There are 100K senders. Each sender's fair share bottleneck bandwidth is 100Kbps.}}
\end{figure}

\paragraph{Strategic Attacks:} Attackers may launch sophisticated
attacks (\S~\ref{sec:security}) than brute-force flooding attacks.  We
simulate microscopic on-off attacks and show that with \sys, the
attack traffic's shape does not reduce a legitimate user's throughput.

The simulation topology is the same as in the previous
single-bottleneck simulations. All legitimate users send long-running
TCP traffic, while attackers send on-off UDP traffic. In the on-period
$T_{on}$, an attacker sends at the rate of 1Mbps; in the off-period
$T_{off}$, it does not send any traffic. All attackers synchronize
their on-periods to create the largest traffic bursts. There are 100K
simulated senders, each having a fair share bandwidth of at least
100Kbps.

In these simulations, we use two different values for $T_{on}$: 0.5s
and 4s. For each $T_{on}$, we vary the off-period length $T_{off}$
from 1.5s to 100s.  Figure~\ref{fig:sim-on-off} shows the simulation
results. As we can see, the average user throughput is at least a
user's fair share rate as if attackers were always active (100Kbps),
indicating that the attack cannot reduce a legitimate user's fair
share of bandwidth. As the attackers' off-period length increases
toward 100s, a legitimate user can achieve a throughput close to
400Kbps, indicating that long running TCP users can use most of the
bottleneck bandwidth when the attackers' off-period is long.

\section{Discussion}
\label{sec:discussion}

\paragraph{Fair Share Bound:} When a disproportionally large
number ($B$) of attackers attack a narrow link $C$ (\eg, when a million
bots attack a 1Mbps link), the fair share lower bound
$O(\frac{C}{G+B})$ achieved by \sys or per-sender fair
queuing (\eg, \cite{stopit}) is small. However, this lower bound
is still valuable, because without it, a small number of attackers can
starve legitimate TCP flows on a well-provisioned link (\eg, 10Gbps).
Although this guarantee does not prevent large-scale DoS attacks from
degrading a user's network service, it mitigates the damage of such
attacks with a predictable fair share, without trusting receivers or
requiring the network to identify and remove malicious traffic. Other
means, like congestion quota discussed below, can be used to further
throttle malicious traffic.

\paragraph{Equal Cost Multiple Path (ECMP):} \sys assumes that a
flow's path is relatively stable and the bottleneck links on the path
do not change rapidly. One practical concern arises as routers may
split traffic among equal-cost multi-paths for load balancing.
Fortunately, most ECMP implementations in practice (\eg,
\cite{cisco-ecmp}) would assign a flow's packets to the same path to
avoid packet reordering. Thus, we expect \sys to work well with ECMP.

\paragraph{Congestion Quota:} If we assume legitimate users have
limited traffic demand at attack time while attackers aim to persistently congest a
bottleneck link, we can further weaken a DoS flooding attack by
imposing a congestion quota, an idea borrowed from
re-ECN~\cite{reecn}. That is, an access router only allows a host to
send a limited amount of ``congestion traffic'' through a bottleneck
link within a period of time. Congestion traffic can be defined as the
traffic that passes a rate limiter when its rate limit decreases.
With a congestion quota, if an attacker keeps flooding a link, its
traffic through the link will be throttled after it consumes its
congestion quota. Different from re-ECN, \sys can impose a
per-(sender, bottleneck link) congestion quota so that a sender's
traffic not traversing any link that is under attack will not be
incidentally throttled when its quota for a link under attack is used
up.

\paragraph{Congestion Detection and Rate Limit Control:} For
simplicity, the current implementation of \sys uses RED as the
congestion detection algorithm, and the \sys design uses a closed-loop
AIMD algorithm to adjust rate limits. RED is not a strictly load-based
congestion detection algorithm. In the future, we plan to explore a
load-based congestion detection algorithm that may react sooner to
congestion than RED. Beside AIMD, another option to set a sender's
rate limit is to let a congested router compute a max-min share rate
of each sender, and periodically send this rate to a sender's access
router. We discard this approach because it involves per-sender state
and has higher computation and message overhead than AIMD.

\paragraph{Control Interval Length:} The current \sys design uses a
fixed control interval $I_{lim}$ for all senders. It is difficult to
optimize the value of $I_{lim}$, because a short $I_{lim}$ cannot
accommodate heterogeneous RTTs on the Internet, while a long $I_{lim}$
slows down the convergence of rate limits. A future improvement we
plan to explore is to have a variety of control interval lengths
$\{I_{lim}^k\}$. A sender may signal which length it prefers to use
with a few bits in its \sys header. Accordingly, we need to update the
rate limit adjustment algorithm on an access router and the congestion
hysteresis on a congested router to ensure fairness among senders with
different control interval lengths.

\paragraph{Convergence Speed:} It may take a relatively long time
(\eg, 100s-200s) for \sys to converge to fairness when a sender's rate
limit is significantly different from its fair share of bottleneck
bandwidth. This is because the control interval $I_{lim}$ is on the
order of a few seconds (2s in our current implementation), much longer
than a typical RTT on the Internet. This convergence speed is
acceptable in the \sys design, because a rate limiter persists for a
much longer period of time (\ie, on the order of hours), which
prevents attackers from constantly taking advantage of the unfairness
during the convergence by sending strategic on-off traffic.

\paragraph{TCP and Rate Limiter Interaction:} TCP's window size
adjustment may be temporarily out of synchronization with the rate
limiter adjustment. For instance, TCP may just receive its ACK before
the rate limit is reduced, in which case TCP's sending rate will
increase. This temporary out of synchronization is not a big issue,
because the rate limit adjustment interval $I_{lim}$ is much larger
than a typical RTT in the Internet such that the out of
synchronization may occur at most once among many RTTs.


\section{Related Work}
\label{sec:related}

At the architectural level, \sys combines the elements of
capability-based systems~\cite{tva-ton,siff,portcullis} and
re-ECN/re-feedback~\cite{refeedback,reecn}. In contrast to capability
tokens, \sys's congestion policing feedback carries valuable network
congestion information. Re-ECN/re-feedback is a congestion policing
framework that incentivizes rational senders to honestly report
downstream path congestion. Routers will discard the packets from the
senders that under-report downstream congestion with high probability
before they reach the destinations.  In contrast, \sys is a DoS
defense architecture that uses unspoofable congestion policing
feedback to scalably and robustly guarantee a sender's fair share of
bottleneck bandwidth in face of attacks. Attackers cannot send packets
with false congestion feedback reporting no or low levels of
congestion to flood a link. Instead, they can at most send packets
reporting the actual levels of congestion and will not gain more
bandwidth than honest senders. In addition, DoS victims can use the
unspoofable feedback as capability tokens to suppress unwanted
traffic. ECN-nonce~\cite{ecn-nonce} robustly signals congestion from
the network to a honest sender even when a receiver attempts to hide
congestion, while \sys enables robust congestion signaling from
congested routers to access routers when both senders and receivers
are malicious.

\sys's request packet protection mechanism is inspired by
Portcullis~\cite{portcullis} that uses computational puzzles to impose
delay on senders. Differently, \sys uses a rate limiting algorithm
that does not require proof-of-work (PoW) nor a network-wide puzzle
synchronization mechanism. This algorithm is similar in spirit to
LazySusan~\cite{lazy-susan} which substitutes resource-based PoW for
latency-based PoW. Different from LazySusan, \sys uses a sender's
waiting time to set its request packet's priority level, and
guarantees the eventual delivery of a legitimate request packet.

Several DoS defense systems aim to enable a victim to install
network filters to stop unwanted traffic~\cite{aitf-ton,aip,stopit},
or to control who can send to it~\cite{off-default}. Unlike them, \sys
does not use per-host queues at congested routers to separate
legitimate traffic from attack traffic in case compromised receivers
collude with malicious senders. Pushback~\cite{pushback} sends
hop-by-hop pushback messages from a congested router to install
per-(incoming interface, destination prefix) rate limiters to reduce
DoS flooding traffic. \sys does not require hop-by-hop deployment,
enables a victim to suppress unwanted traffic, and provides per-sender
fairness at bottleneck links: attackers cannot diffuse their traffic
to many destinations to gain unfair bandwidth shares. AIP~\cite{aip}
uses trusted host hardware to block unwanted attack traffic, while
\sys places policing functions inside the network and does not require
trusted host hardware.

Speakup~\cite{defense-by-offense} and Kill-Bots~\cite{dina-bots}
address application-layer DoS attacks, while \sys addresses
network-layer DoS attacks.  Several systems use overlay
networks~\cite{mayday, sos, dixon2008, stavrou-ccs-2005, overdose,
centertrack} or middleboxes~\cite{flow-cookies, dfense} to mitigate
DoS attacks against dedicated destinations. DoS mitigation products on
today's market (\eg, \cite{stratecast-2010}) offer in-network anomaly
detection and attack traffic removal services near the victims.
Kreibich et al.~\cite{packet-symmetry} propose to use packet symmetry
to detect and remove attack traffic.  This body of work requires fewer
changes to routers, but \sys can remove attack traffic near its
origins and protect all destinations on the Internet once deployed.
Moreover, it places the attack traffic identification function at the
receivers to keep the network open to new applications.

\sys's approach to scalability is inspired by
CSFQ~\cite{core-stateless} that achieves per-flow fairness without
per-flow queues in the core routers. Differently, \sys enables DoS
victims to suppress attack traffic, and provides per-sender rather
than per-flow fairness.

\section{Conclusion}
\label{sec:conclusion}

This paper presents the design and evaluation of \sys, an architecture
that places the network at the first line of DoS defense. \sys uses a
key mechanism, secure congestion policing feedback, to enable scalable
and robust traffic policing inside the network. Bottleneck routers use
the congestion policing feedback to signal congestion to access
routers, and access routers use it to robustly police senders'
traffic.  In case compromised senders and receivers collude in pairs
to flood the network, \sys limits the damage of this attack by
providing each sender (malicious or legitimate) its fair share of
bottleneck capacity without keeping per-host state at bottleneck
routers. In case attackers send DoS floods to innocent victims, \sys
enables the DoS victims to use the secure congestion policing feedback
as capability tokens to suppress unwanted traffic. Using a combination
of a Linux implementation, simulations, and theoretic analysis, we
show that \sys is an effective DoS solution that reduces the amount of
state maintained by a congested router from
per-host~\cite{tva-ton,stopit} to per-AS.

\section*{Acknowledgment}

The authors thank Jeff Chase, David Harrison, Yongqiang Liu, and the
anonymous SIGCOMM reviewers for their insightful comments, and David
Oran for shepherding this paper. This work is supported in part by NSF
awards CNS-0925472 and CNS-0845858.

{\small \bibliographystyle{abbrv} \bibliography{bibtex/ddos}}

\appendix

\section{Convergence Analysis}
\label{sec:appendix_proof}

NetFence uses an AIMD algorithm to adjust a sender's rate limit. AIMD
has been proven to converge onto efficiency and fairness~\cite{aimd}.
We first briefly summarize the AIMD results, and then analyze how well
NetFence converges to fairness and efficiency.

\subsection{AIMD Preliminary}
\label{sec:app_aimd}

Consider a simplified fluid model where one link is shared by $N$ synchronous flows, all of whom having the same round-trip time (RTT), $T$. Each flow $i$ uses the following rule to update its sending rate $x_i(t)$ at time $t$:

\vspace{0.03in}
\hspace{0.2in} AI: \ \ \ $x_i(t+T) \ = \ x_i(t) + \alpha$  \ \ \ \ \ if no congestion

\hspace{0.2in} MD: \ $x_i(t+\delta t) \ = \ x_i(t) \times \beta$ \ \ \ \ \ otherwise

\vspace{0.03in}
\hspace{-0.12in}where $\alpha > 0$, $1 > \beta > 0$, $\delta t > 0$ and $\delta t \rightarrow 0$.

\vspace{0.02in}
{\em Convergence to efficiency}: When the link is under-utilized, AI is applied. The aggregate rate continues to increase as $\sum_i x_i(t+T) = \sum_i x_i(t) + N \alpha > \sum_i x_i(t)$. At some point the link will eventually be over-utilized, leading to $\sum_i x_i(t+\delta t) = \beta \times \sum_i x_i(t) < \sum_i x_i(t)$, a cut of the aggregate rate. MD finally results in link under-utilization. This pattern of oscillation around full utilization repeats, with the overshoot and undershoot decided by $N\alpha$ and $\beta$, respectively. Efficiency convergence is achieved if $\alpha \rightarrow 0$ and $\beta \rightarrow 1$.

{\em Convergence to fairness}: We measure fairness with Jain's
fairness index~\cite{aimd}: $(\sum_i x_i(t))^2 / N \sum_i x_i^2(t)$.
It is easy to show that MD keeps the index unchanged, while AI
increases it. So, given sufficient number of AI rounds, the fairness
keeps increasing, and will eventually reach its maximum, 1. At this
point, all the flows share the link capacity equally.

\subsection{Capacity Share Lower Bound}
\label{sec:app_bound}

Given $G$ good and $B$ bad hosts sharing a bottleneck link of capacity
$C$, can NetFence guarantee that each good host obtains its fair share
$O(\frac{C}{G+B})$ of the capacity, regardless of the attack strategy
bad hosts may apply? Here we prove this is indeed true.

\begin{figure}
\centering
\includegraphics[height=0.9in, width=3.1in]{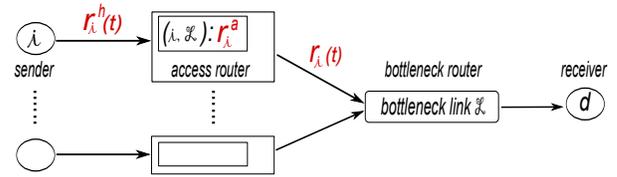}
\caption{\small A simplified fluid model for NetFence}
\label{fig:model}
\end{figure}

Again we consider a simplified fluid model in Figure~\ref{fig:model}.
There are a fixed number of senders and one receiver. To make our
analysis tractable, we do not consider delay: any rate change at the
senders immediately takes effect at the bottleneck link. The bad hosts
can use traffic of any shape to attack. As shown in the figure, for
any host $i$, let $r_i^h(t)$ be its sending rate at time $t$, and
$r_i^a$ be its rate limit at its access router. $r_i^a$ is a constant
value during a control interval [$t_0, \, t_0+I_{lim}$), where $t_0$
is a particular point in time. Then, at the output link of the access
router, the egress rate of the host $i$ is $r_i(t)$ = min($r_i^h(t)$,
$r_i^a$) due to the rate limit.

\vspace{0.05in}
{\em \textbf{Definition}: We say a host has \textbf{sufficient demand} from its application if the host sends fast enough such that its corresponding rate limit is not punished by the rate limit AIMD algorithm described in \S~\ref{sec:robust-rate-probing}.}

\vspace{0.05in}
{\em \textbf{Definition}: During any control interval, given a rate limit $r_i^a$, the \textbf{rate limit utilization} $\nu_i = \overline{r_i} / r_i^a$, where the host average egress rate $\overline{r_i} = \frac{1}{I_{lim}} \int_{t_0}^{t_0+I_{lim}} r_i(t) dt$. Obviously, we have $0 < \nu_i \le 1$.}

\vspace{0.03in}
Remark: If a sender $i$'s traffic is sent in TCP, $\nu_i$ depends on
TCP's congestion control efficiency. Given enough buffers at the
congested router, this $\nu_i$ is close to 1 in typical scenarios,
because NetFence uses a control interval $I_{lim} = 2s$ for rate limit
AIMD, which is typically one order of magnitude larger than the
end-to-end RTT (on average 100$\sim$200ms in the Internet). We might
have a low $\nu_i$ (\eg, less than 50\%) under very low/high
bandwidth-delay product networks where TCP is inefficient. If a sender
$i$'s traffic is sent in UDP, $\nu_i$ depends on the host sending rate
$r_i^h(t)$ during the control interval $I_{lim}$. For a source that
sends as fast as allowed by the rate limiter, we have $\nu_i = 1$; for
on-off traffic, $\nu_i$ is decided by its duty cycle.

\vspace{0.05in}
{\em \textbf{Lemma}: For any host $s$ with sufficient demand, its rate limit will eventually be the highest among all the rate limits, \ie, $r_s^a = \texttt{max}_i(r_i^a)$.}

\vspace{0.03in}
Proof. We first show that if all the hosts have sufficient demand,
they will eventually reach the same $r_i^a$. This is due to the AIMD
rule. As we have shown in Section~\ref{sec:app_aimd}, MD keeps
fairness unchanged, but AI increases it. Under NetFence, after an MD,
there is at least one AI round. This is because if there were only MD
rounds, the aggregated rate limit would eventually become very small,
and then the bottleneck link would not be congested, which would lead
to an AI. Therefore, the rate limit fairness increases as time goes
on, and eventually all the hosts will have the same $r_i^a$. For a
host without sufficient demand, NetFence punishes it by not increasing
or even decreasing its rate limit. Therefore, the host can only get a
lower rate limit than one with sufficient demand. $\Box$

\vspace{0.05in}
{\em \textbf{Assumption}: Congestion detection signals if and only if aggregate demand surpasses bottleneck link capacity, \ie, when $\sum_i \overline{r_i} \ge C$.}

\vspace{0.03in}
Remark: In practice, this is often true for load-based congestion
detection schemes that signal congestion when the average traffic
arrival rate exceeds a high-load threshold. Queue-based congestion
detection (like RED) is more sensitive to traffic dynamics, possibly
signaling congestion due to bursty traffic even if the link average
utilization in one control interval is low. However, in the case of
NetFence, according to our simulations, queue-based congestion
detection satisfies this assumption well. This is because each
sender's traffic is shaped by a rate limiter before it enters a
bottleneck, which significantly limits the peak rate of the aggregate
incoming traffic.

\vspace{0.05in}
{\em \textbf{Theorem}: Given $G$ good and $B$ bad senders sharing a bottleneck link of capacity $C$, regardless of attack strategy, any good sender $g$ with sufficient demand eventually obtains a fair share $\frac{\nu_g \rho C}{G+B}$ where $\rho = (1-\delta)^3$.}

\vspace{0.03in}
Proof. During a congested control interval, we have $C \le \sum_i
\overline{r_i} \le \sum_i r_i^a \le \sum_i \texttt{max}_i(r_i^a) =
r_g^a (G+B)$ and thus $r_g^a \ge \frac{C}{G+B}$. For any uncongested
control interval, the rate limit is reduced from the peak rate at the
congested control interval, so $r_g^a \ge \frac{\rho C}{G+B}$ where
$\rho = (1-\delta)^3$. This particular $\rho$ value is due to the fact
that our rate limit adjustment may apply one MD cut to make the link
not congested and then at most two more extra MD cuts to prevent
malicious senders from hiding \ldown feedbacks. Therefore, we have,
for all the control intervals in the steady state, $\overline{r_g} =
\nu_g r_g^a \ge \frac{\nu_g \rho C}{G+B}$. $\Box$

\vspace{0.03in}
Remark: For a multi-homed host $h_m$ that uses the bottleneck link $l$ via multiple access routers, there is one rate limiter ($h_m$, $l$) in {\em each} access router. The host's bandwidth share can thus be higher than a single-homed host. This is not a big issue since the number of access routers of any multi-homed host is often small (\eg, 2). It is also reasonable since the host pays for multiple access links.

\section{Alternative Designs to Handle Multiple Bottlenecks}
\label{sec:appendix_mbtnk}

Next we present two possible solutions to improve \sys's
performance in the multiple bottleneck case as described in
\S~\ref{sec:multi-ratelimiters}.

\subsection{Multi-bottleneck Feedback in One Packet}
\label{sec:app-multi-feedback}

A clean solution to handle multiple bottlenecks is to let a
single packet carry the congestion policing feedback from multiple
bottleneck links on a forwarding path. Compared to \sys's core design
described in \S~\ref{sec:design}, this solution requires a new \sys
header format, a new set of algorithms to stamp and verify feedback,
and a new regular packet policing algorithm; other parts of \sys's
core design do not need to change.

\paragraph{Multi-bottleneck feedback in a \sys\ header:} A \sys header
may carry the congestion policing feedback from zero or more
bottleneck links on the packet's forwarding path. A link $L$ may stamp
one of two types of feedback: \lup\ or \ldown. Their meanings are the
same as defined in \S~\ref{sec:feedback}. The $i$th bottleneck's
feedback in a packet is encoded in two fields: $link_i$ and
$action_i$, whose values are the same as defined in
\S~\ref{sec:secure_feedback}.

All the feedback fields in a \sys header are made unforgeable by a
single $token$ field, as we will show soon. A \sys header also
contains a single timestamp field $ts$ to indicate the freshness of
all the feedback.

If a packet does not traverse any bottleneck link, its \sys header
will only contain a $ts$ field and a valid $token$ field stamped by
its access router. We refer to such a \sys header as containing the
\ok feedback.

\begin{figure}[t]
\centering
\includegraphics[width=0.9\columnwidth]{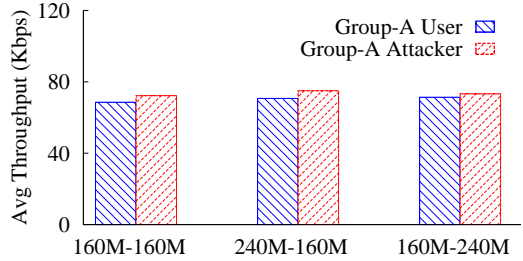}
\caption{\label{fig:multi-data-flooding-multifb} {\bf \small Sender
throughput (Kbps) with the same simulation setting as in
Figure~\ref{fig:multi-data-flooding}, but with each packet carrying
congestion policing feedback from multiple bottleneck links. The fair
share rate for each sender is 80Kbps.}}
\end{figure}

\paragraph{Stamping and verifying feedback:} When a packet leaves its
access router, the access router stamps the \ok feedback. That is, it
sets $ts$ to its local time and computes the $token$ field as follows:
\begin{equation}
token_{nop} = MAC_{K_a}(src, dst, ts) \label{eq:nop-multi}
\end{equation}
$K_a$ is the same as defined in \S~\ref{sec:secure_feedback}.

When the packet traverses a bottleneck link $L$ in \bad\ state, the
bottleneck router inserts its feedback into the \sys header. It
inserts \ldown if the algorithm in
\S~\ref{sec:robust-rate-probing} determines to do so, or \lup
otherwise. The bottleneck router updates the $token$ field as follows:
\begin{equation}
token = MAC_{K_{ai}}(src, dst, ts, L, action, token) \label{eq:feedback-multi}
\end{equation}
$K_{ai}$ is the same as defined in \S~\ref{sec:secure_feedback}. The
computation of the new $token$ value covers the old $token$ value to
prevent a malicious downstream router from tampering the feedback
stamped by other bottleneck links.

When an access router verifies a \sys header, it first checks the $ts$
field as described in \S~\ref{sec:secure_feedback}. Then it
reconstructs the original $token_{nop}$ using Eq~(\ref{eq:nop-multi}),
and recomputes the $token$ field using Eq~(\ref{eq:feedback-multi}).
When the packet carries feedback from multiple bottlenecks, the router
will have to apply Eq~(\ref{eq:feedback-multi}) multiple times to
calculate the final $token$ value.

\paragraph{Policing regular packets:} The multi-bottleneck feedback in
a regular packet clearly indicates the bottleneck links on the
packet's forwarding path. An access router rate-limits a regular
packet with all the rate limiters each associated with one on-path
bottleneck link. If the packet cannot pass any of them, it is
discarded. This algorithm prevents the rate limit for a flow from
changing abruptly, and it also ensures that the flow's sending rate is
smaller than the lowest rate limit for all the on-path bottleneck
links.

\paragraph{Simulation results:}
Figure~\ref{fig:multi-data-flooding-multifb} shows the simulation
results when a packet can carry multi-bottleneck feedback. The simulation
setting is the same as in Figure~\ref{fig:multi-data-flooding}. As can
be seen, each sender in Group-A can get roughly its fair share of
bottleneck bandwidth.

\begin{figure}[t]
\centering
\includegraphics[width=0.9\columnwidth]{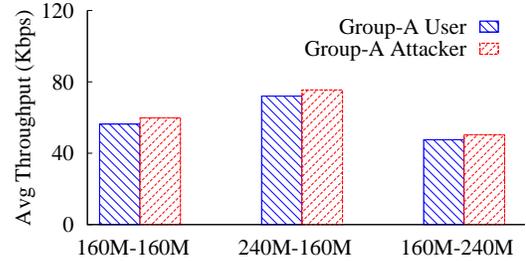}
\caption{\label{fig:multi-data-flooding-rlcache} {\bf \small Sender
throughput (Kbps) with the same simulation setting as in
Figure~\ref{fig:multi-data-flooding} but also with rate limiter
inference. The fair share rate for each sender is 80Kbps.}}
\end{figure}

\subsection{Inferring Rate Limiters}
\label{sec:app-infer-rate-limiter}

The solution in Appendix~\ref{sec:app-multi-feedback} requires a new
\sys\ header format, which may be longer than that in the core \sys\
design described in \S~\ref{sec:design}. An alternative design is that
a packet still carries the congestion policing feedback from a single
bottleneck, but an access router can infer the bottleneck links a flow
traverses and police the flow's packets using all the corresponding
rate limiters. This solution is compatible with \sys's core design,
because each access router can independently deploy it.

\paragraph{Inferring on-path rate limiters:} An access router keeps a
per-destination-prefix cache that records what bottleneck links are on
the path towards a particular prefix. Whenever an access router
receives \lup or \ldown feedback from a sender, it adds $L$ into the
cache entry associated with the destination prefix. $L$ may be removed
from the cache entry if all the rate limiters for $L$ have been
removed or the \lup and \ldown feedback has not appeared in packets
toward the destination prefix for a long period of time.

The number of entries in the inference cache is at most the number of
prefixes in a full BGP table. An access router also needs the prefix
list of a full BGP table in order to locate the cache entry given the
destination IP address in a packet.

\paragraph{Policing regular packets:} Based on the inference cache, an
access router passes a packet with \bad feedback through all the rate
limiters associated with the bottleneck links on the packet's
forwarding path. If the packet cannot pass any of them, it is
discarded.

The inference cache may be inaccurate: a link $L$ may stay in the
cache entry of a particular prefix even after it is no longer in \bad
state or on the path toward the prefix. In this case, traffic toward
the destination prefix may be unnecessarily throttled by the rate
limiters associated with $L$. However,this is not a big problem
because such cases only occur infrequently, and $L$ will be removed
from the cache entry after traffic toward the destination prefix no
longer carries the \lup and \ldown feedback for some time.

When a packet with the \lup or \ldown feedback passes all the
associated rate limiters, let $(src,L_{low})$ denote the one with the
smallest rate limit. Unlike the algorithm described in
\S~\ref{sec:police-regular}, the access router will reset the feedback
to $L_{low}^{\uparrow}$.

\paragraph{Inferring rate limits:} With the rate limiter inference
cache, the single-bottleneck feedback in a packet may be used to infer
the feedback from other bottleneck links on the packet's forwarding
path.  If a packet carries the \lup feedback, other on-path bottleneck
links must not be congested at this moment, because otherwise the \lup
feedback would have been replaced with an $L^{* \downarrow}$ feedback;
on the other hand, if a packet carries an \ldown feedback, other
on-path bottleneck links will not be able to stamp their feedbacks,
and therefore the rate limits for them should be temporarily held
unchanged.

With this inference algorithm, the rate limit adjustment algorithm in
\S~\ref{sec:robust-rate-probing} needs to be updated accordingly. In
addition to $t_s$ and $hasIncr$, a rate limiter $(src, L)$ is also
associated with three more state variables: $hasIncr*$, $isActive$,
and $isActive*$. $hasIncr*$ records whether the rate limiter has seen
the $L^{* \uparrow}$ feedback with a timestamp newer than $t_s$;
$isActive$ records whether the rate limiter has seen any \lup\ or
\ldown\ feedback regardless of the timestamp value; and $isActive*$
records whether the rate limiter has seen any $L^{* \uparrow}$ or
$L^{* \downarrow}$ feedback. At the end of each control interval
$I_{lim}$, the rate limiter $(src, L)$'s rate limit $r_{lim}$ is
adjusted as follows:

\begin{enumerate*}
\item If $hasIncr$ or $hasIncr*$ is true, the rate limiter's
throughput $r_{tput}$ is compared with $\frac{1}{2} r_{lim}$:
$r_{lim}$ is increased by $\Delta$ if $r_{tput} \ge r_{lim}$;
\item Otherwise, if $isActive$ is true, $r_{lim}$ is decreased to
$(1-\delta) r_{lim}$;
\item Otherwise, if $isActive*$ is true, $r_{lim}$ is kept unchanged;
\item Otherwise, $r_{lim}$ is decreased to $(1-\delta) r_{lim}$.
\end{enumerate*}

\paragraph{Simulation results:}
Figure~\ref{fig:multi-data-flooding-rlcache} shows the simulation
results with the above rate limiter inference algorithm. The
simulation setting is the same as in
Figure~\ref{fig:multi-data-flooding} and
Figure~\ref{fig:multi-data-flooding-multifb}. We can see that with
this algorithm, TCP senders and attackers in Group-A have roughly the
same throughput, because the rate limit of a flow no longer jumps
between significantly different values. The throughput of a Group-A
sender is improved compared to Figure~\ref{fig:multi-data-flooding};
however, it may still be much smaller than its fair share. This is
because for any Group-A sender $src$, the rate limits of both rate
limiters $(src,L_1)$ and $(src, L_2)$ can only increase when neither
bottleneck links is congested. This is a fundamental limitation of
embedding single-bottleneck feedback in a packet: the packet simply
cannot carry enough congestion information.

\section{Pseudo Code}

This section includes the pseudo code of \sys's key procedures.

\begin{figure}[t]
\centering {\footnotesize
\begin{algorithmic}[1]
\Procedure{rate\_limiter.rate\_limit\_request\_packet}{$pkt$}
\State $hdr$ $\gets$ get\_netfence\_header($pkt$)
\If {$hdr.priority==0$}
  \State \textbf{return} $PASS$
\EndIf
\State $ts_{now}$ $\gets$ get\_current\_time()
\State $token_{now}$ $\gets$ $m\_token_{init}+(ts_{now}-m\_ts_{init})*m\_rate_{init}$
\State $token_{remove}$ $\gets$ $2^{hdr.priority-1}$
\If {$token_{remove}>token_{now}$}
  \State \textbf{return} $DROP$
\EndIf
\If {$token_{now}>m\_depth_{init}$}
  \State $token_{now}$ $\gets$ $m\_depth_{init}$
\EndIf
\State $m\_token_{init}$ $\gets$ $token_{now}-token_{remove}$
\If {$m\_token_{init}<0$}
  \State $m\_token_{init}$ $\gets$ $0$
\EndIf
\State $m\_ts_{init}$ $\gets$ $ts_{now}$
\State \textbf{return} $PASS$
\EndProcedure
\end{algorithmic}}
\caption{\label{fig:code-request-limiting} {\bf \small \sys request packet rate limiting pseudo-code.
$m\_*$ are member variables of the rate limiter.}}
\end{figure}

\begin{figure}[t]
\centering {\footnotesize
\begin{algorithmic}[1]
\Procedure{rate\_limiter.rate\_limit\_regular\_packet}{$pkt$}
\State rate\_limiter.update\_outgoing\_rate($pkt.len$)
\State $r$ $\gets$ rate\_limiter.cache\_packet($pkt$)
\If {$r==CACHED$}
  \State \textbf{return} $CACHED$
\ElsIf {$r==DROP$}
  \State \textbf{return} $DROP$
\EndIf
\State \textbf{return} $PASS$
\EndProcedure
\Procedure{rate\_limiter.cache\_packet}{$pkt$}
\State $ts_{now}$ $\gets$ get\_current\_time()
\If {$m\_cache$.size\_pkts()$==0$}
  \If {$(ts_{now}-m\_ts\_depart_{regular})*m\_rate_{regular} \ge pkt.len*8$}
    \State $m\_ts\_depart_{regular} \gets ts_{now}$
    \State \textbf{return} $PASS$
  \ElsIf {caching\_delay\_too\_long($pkt$)}
    \State \textbf{return} $DROP$
  \EndIf
\EndIf
\State $m\_cache$.enque($pkt$)
\If {$m\_cache$.size\_pkts()$==1$}
  \State rate\_limiter.schedule\_next\_unleash()
\EndIf
\State \textbf{return} $CACHED$
\EndProcedure
\Procedure{rate\_limiter.unleash\_packet}{ }
\If {$m\_cache$.size\_pkts()$==0$}
  \State \textbf{return} $NULL$
\EndIf
\State $pkt$ $\gets$ $m\_cache$.deque()
\State $m\_ts\_depart_{regular}$ $\gets$ get\_current\_time()
\If {$m\_cache$.size\_pkts()$>0$}
  \State rate\_limiter.schedule\_next\_unleash()
\EndIf
\State \textbf{return} $pkt$
\EndProcedure
\end{algorithmic}}
\caption{\label{fig:code-regular-limiting} {\bf \small \sys regular packet rate limiting pseudo-code.
$m\_*$ are member variables of the rate limiter.}}
\end{figure}

\begin{figure}[t]
\centering {\footnotesize
\begin{algorithmic}[1]
\Procedure{rate\_limiter.update\_status}{$pkt$}
\State $hdr$ $\gets$ get\_netfence\_header($pkt$)
\State $ts_{pkt}$ $\gets$ recover\_timestamp\_from\_packet($hdr$)
\If {$ts_{pkt} \ge m\_t_s$}
  \If {$hdr.action==INCR$}
    \State $m\_hasIncr \gets TRUE$
  \EndIf
\EndIf
\EndProcedure
\Procedure{rate\_limiter.adjust\_rate\_limit}{ }
\State $action \gets KEEP$
\State $rate_{old} \gets m\_rate_{regular}$
\If {$m\_hasIncr$}
  \If {rate\_limiter.get\_outgoing\_rate()$>m\_rate_{regular}/2$}
    \State $action \gets INCREASE$
  \EndIf
\Else
  \State $action \gets DECREASE$
\EndIf
\If {$action==INCREASE$}
  \State $m\_rate_{regular} \gets m\_rate_{regular}+\Delta$
\ElsIf {$action==DECREASE$}
  \State $m\_rate_{regular} \gets m\_rate_{regular}*(1-\delta)$
\EndIf
\If {$action \ne KEEP$}
  \State rate\_limiter.update\_packet\_cache($rate_{old}$)
\EndIf
\State $m\_hasIncr \gets FALSE$
\State $m\_t_s$ $\gets$ get\_current\_time\_in\_seconds()
\State rate\_limiter.schedule\_next\_rate\_adjustment()
\EndProcedure
\end{algorithmic}}
\caption{\label{fig:code-rate-adjustment} {\bf \small \sys rate limit adjustment pseudo-code.
$m\_*$ are member variables of the rate limiter.}}
\end{figure}

\begin{figure}[t]
\centering {\footnotesize
\begin{algorithmic}[1]
\Procedure{router.forward\_packet}{$pkt$}
\State $q$ $\gets$ router.find\_output\_queue($pkt$)
\If {$q==NULL$}
  \State discard\_packet($pkt$)
  \State \textbf{return}
\EndIf
\If {is\_legacy\_packet($pkt$)}
  \State $q$.enque($pkt$)
  \State \textbf{return}
\EndIf
\If {router.is\_from\_my\_hosts($pkt$)}
  \State $r$ $\gets$ router.rate\_limit\_packet($pkt$)
  \If {$r==PASS$}
    \State $q$.enque($pkt$)
  \ElsIf {$r==DROP$}
    \State discard\_packet($pkt$)
  \EndIf
\EndIf
\EndProcedure
\Procedure{router.rate\_limit\_packet}{$pkt$}
\State $hdr$ $\gets$ get\_netfence\_header($pkt$)
\If {$hdr.mac==0$ \textbf{or not} router.mac\_is\_valid($pkt$)}
  \State $r$ $\gets$ router.get\_init\_pkt\_rate\_limiter($pkt$)
  \If {$r$.rate\_limit\_init\_packet($pkt$)$==DROP$}
    \State \textbf{return} $DROP$
  \EndIf
\Else
  \If {$hdr.linkid \ne 0$}
    \State $r$ $\gets$ router.get\_regular\_pkt\_rate\_limiter($pkt$)
    \State $r$.update\_status($pkt$)
    \State $x$ $\gets$ $r$.rate\_limit\_regular\_packet($pkt$)
    \If {$x==DROP$}
      \State \textbf{return} $DROP$
    \ElsIf {$x==CACHED$}
      \State \textbf{return} $CACHED$
    \EndIf
  \EndIf
\EndIf
\State router.update\_packet($pkt$)
\State \textbf{return} $PASS$
\EndProcedure
\end{algorithmic}}
\caption{\label{fig:code-forwarding} {\bf \small \sys packet forwarding pseudo-code.}}
\end{figure}

\begin{figure}[t]
\centering {\footnotesize
\begin{algorithmic}[1]
\Procedure{queue.deque\_regular\_packet}{ }
\State $pkt$ $\gets$ queue.pick\_next\_regular\_packet\_to\_deque()
\State $q$ $\gets$ queue.get\_regular\_queue($pkt$)
\If {$q.ts_{mon} \ge 0$}
  \State queue.update\_netfence\_header($pkt$)
\EndIf
\State \textbf{return} $pkt$
\EndProcedure
\Procedure{queue.check\_packet\_loss}{ }
\State $ts_{now}$ $\gets$ get\_current\_time()
\State $qset$ $\gets$ queue.get\_regular\_queue\_set()
\For {$q$ in $qset$}
  \State $dr$ $\gets$ $q$.get\_drop\_rate()/$q$.get\_deque\_rate()
  \State $q.drop\_rate \gets q.drop\_rate*0.9+dr*0.1$
  \If {$q.drop\_rate>p_{th}$}
    \State $q.ts_{mon} \gets ts_{now}$
  \ElsIf {$q.ts_{mon}>0$ \textbf{and} $ts_{now}-q.ts_{mon}>T_{recover}$}
    \State $q.ts_{mon} \gets -1$
  \EndIf
\EndFor
\EndProcedure
\end{algorithmic}}
\caption{\label{fig:code-attack-detection} {\bf \small Pseudo-code
showing how a bottleneck router may update a packet's congestion
policing feedback.}}
\end{figure}

\end{document}